\documentclass[twocolumn]{aastex631}

\shorttitle{The Population of Radio Hypernebulae}
\shortauthors{Uno et al.}

\defcitealias{Sridhar2022ApJ}{SM22}

\graphicspath{{./}{figures/}}
 
\usepackage{amsmath}
\usepackage{xcolor}

\begin{document}

\title{The Population of Radio Hypernebulae from Mass-Transferring Massive Binaries}

\correspondingauthor{Kohki Uno}
\email{ku2204@columbia.edu}

\author[0000-0002-6765-8988]{Kohki Uno}
\affiliation{Department of Astronomy and Columbia Astrophysics Laboratory, Columbia University, New York, NY 10027, USA}

\author[0000-0002-4670-7509]{Brian D.~Metzger}
\affiliation{Department of Physics and Columbia Astrophysics Laboratory, Columbia University, New York, NY 10027, USA}
\affiliation{Center for Computational Astrophysics, Flatiron Institute, 162 5th Avenue, New York, NY 10010, USA}

\author[0000-0002-5519-9550]{Navin Sridhar}
\affiliation{Department of Physics, Stanford University, 382 Via Pueblo Mall, Stanford, CA 94305, USA} 
\affiliation{Kavli Institute for Particle Astrophysics \& Cosmology, 452 Lomita Mall, Stanford, CA 94305, USA} 
  
\author[0000-0001-5484-4987]{Lieke A. C. van Son}
\affiliation{Center for Computational Astrophysics, Flatiron Institute, 162 5th Avenue, New York, NY 10010, USA}
\affiliation{Department of Astrophysical Sciences, Princeton University, Princeton, NJ 08544, USA} 

\author[0000-0002-4421-4962]{Floor S.~Broekgaarden}
\affiliation{Department of Astronomy and Astrophysics, University of California, San Diego, La Jolla, CA 92093, USA}

\author[0000-0001-9584-2531]{Dillon Z. Dong}
\affiliation{National Radio Astronomy Observatory, 1003 Lopezville Road, Socorro, NM 87801, USA}

\author[0000-0003-2611-7269]{Keiichi Maeda}
\affiliation{Department of Astronomy, Kyoto University, Kitashirakawa-Oiwake-cho, Sakyo-ku, Kyoto, 606-8502, Japan}

\begin{abstract}

Massive binary evolution is punctuated by brief phases of highly super-Eddington mass transfer (MT) onto neutron stars and black holes, presaging luminous transients such as supernovae, common envelope events, and gravitational wave sources. These hyper-accreting phases power relativistic jets and slower mass-loaded outflows, inflating extended nebulae of relativistic particles. Though observed in the local universe as ultra-luminous X-ray binaries and micro-quasars, the present census is biased against shorter-lived and hence rarer but potentially even more extreme MT systems: so-called ``hypernebulae'' (HNbe). We use the binary population synthesis code (\textsc{COMPAS}) to estimate the MT population, which spans a wide range of MT rates $\dot{M} \sim 10^{2-7}\,\dot{M}_{\rm Edd}$ and source lifetimes $t_{\rm acc} \sim 10^{0-6}$ years. Combining this synthetic population with a semi-analytic model for their time-dependent radio emission, we predict the observable properties of HNbe over cosmic time. HNbe exhibit a broad GHz luminosity range extending to $L_{\nu, \rm peak} \sim 10^{28-30}\,\mathrm{erg\,s^{-1}\,Hz^{-1}}$, and up to $\sim 10^{2}$ ($\sim 10^{5}$) HNbe are detectable by the wide-field radio surveys VLASS (DSA-2000) in our conservative model. A fraction of these systems should possess angular sizes and spatial offsets from star-forming regions (due to the first supernova kick) resolvable by VLBI. The most luminous donors, helium Hertzsprung-gap stars, may be detectable coincident with radio HNbe with \textit{Roman} imaging out to distances $\sim 30$\,Mpc. Our findings suggest HNbe as promising probes of the pathways linking massive binaries to energetic transients.

\end{abstract}


\section{Introduction} \label{sec:introduction}

\begin{figure}
\centering
\plotone{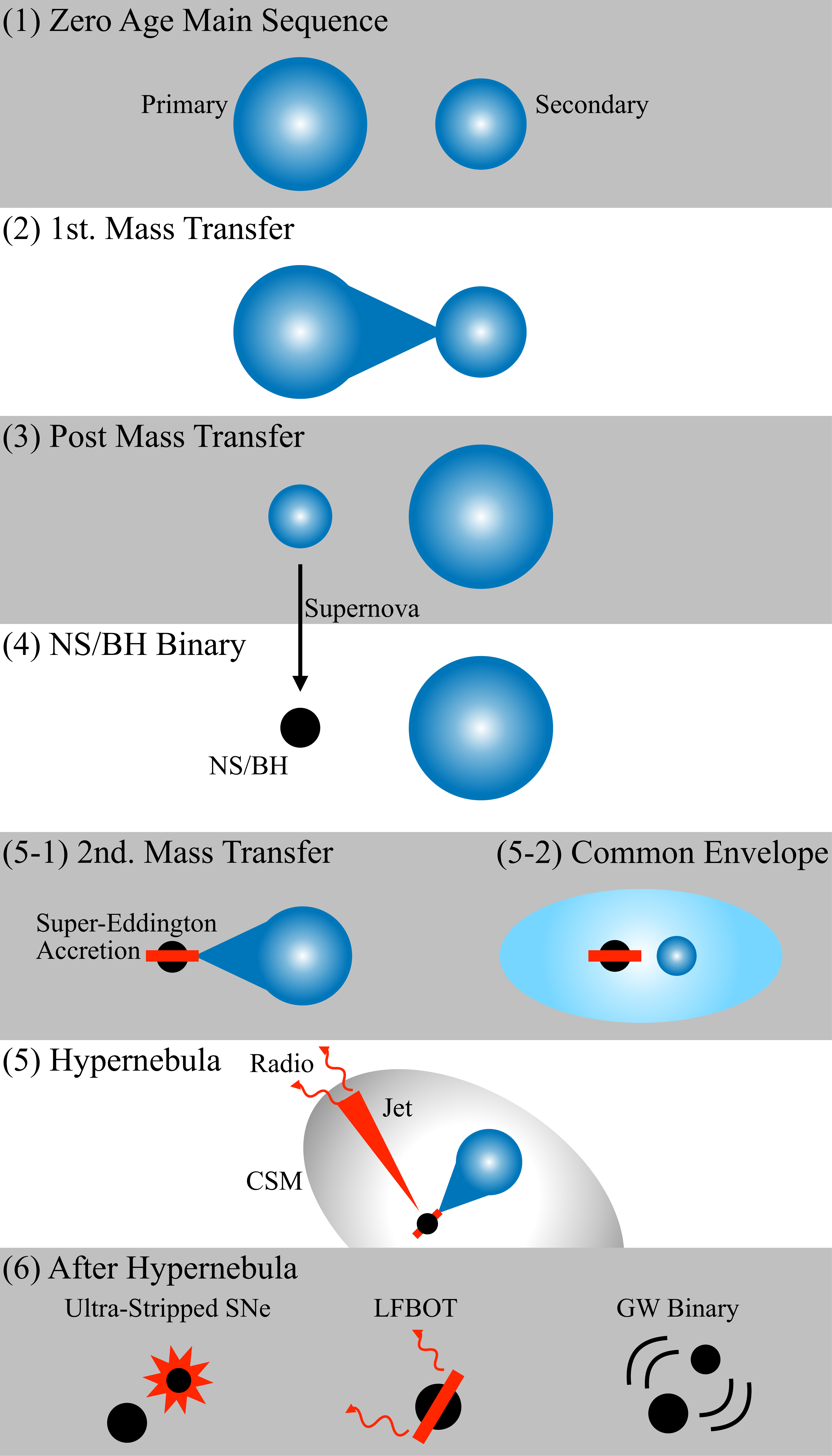}
\caption{Schematic illustration of binary evolution pathways leading to highly super-Eddington mass-transfer phases and the creation of hypernebulae. Following the hypernebula phase, some systems may evolve to generate energetic transients, including ultra-stripped supernovae, tidal disruption events of massive helium stars (possibly generating luminous fast blue optical transients), and compact-object binaries detectable as gravitational wave events. 
}
\label{fig:schematic}
\end{figure}

Massive star binaries across cosmic time play a key role in shaping the Universe across a wide variety of astrophysical scales, epochs and processes, including feedback in galaxies (e.g., \citealt{Grudic+22}), chemical enrichment (e.g., \citealt{Matteucci21,Heger+23}), and reionization of the early Universe (e.g., \citealt{Gotberg+20}). Their evolution also determines the formation channels for some of the most energetic transients in the universe, including supernovae, gamma-ray bursts, and compact object binaries observed through  gravitational waves (e.g., \citealt{Mandel&Broekgaarden22,Marchant&Bodensteiner24}).   
Understanding the physical processes governing binary evolution is therefore essential for connecting stellar evolution to these diverse astrophysical phenomena.

A defining feature of binary evolution, absent in single stars, is the exchange of mass and angular momentum between the stellar components. Mass transfer (MT), driven either by wind-fed accretion or Roche-lobe overflow (e.g., \citealt{Sana+12,Offner+23}), can occur on a wide range of timescales depending on the evolutionary state of the donor star and the binary mass ratio \citep{Pavlovskii+17,vanSon+22,Klencki+22}. In some cases, mass transfer becomes dynamically unstable -- typically when the donor expands in response to mass loss faster than its Roche lobe can adjust -- leading to runaway mass loss and the onset of a common-envelope (CE) phase \citep{Ivanova+13,Ge+20}. These phases are key drivers of binary orbital evolution and can power luminous transients such as luminous red novae (e.g., \citealt{Tylenda+11,Pejcha+17,Matsumoto&Metzger22}) and, potentially, superluminous supernovae (e.g., \citealt{Chevalier_12,Schroder+20}) and luminous fast blue optical transients (LFBOTs; e.g., \citealt{Soker+19,Metzger22,Uno&Maeda23,Klencki&Metzger26}).

When the accretor is a compact object, particularly a black hole (BH) or neutron star (NS), the accretion becomes an additional and highly efficient energy source \citep{Frank+02}. Such systems can be observed as high-mass X-ray binaries in the Milky Way and nearby galaxies \citep[e.g., ][]{Kretschmar+19}. Beyond their intrinsic interest, they are widely considered progenitors or evolutionary intermediates of compact-object binaries that later merge as gravitational-wave sources \citep{LIGO+16,LIGO+19CATALOG,Belczynski+24}. The accretion rate into these systems can approach or exceed the Eddington limit; giving rise to ``ultra-luminous X-ray sources'' (ULXs), for which the apparent isotropic X-ray luminosity of face-on systems can greatly exceed the Eddington luminosity \citep{Kaaret+17,King+23}.

Super-Eddington accretion is inherently accompanied by strong mass loss through disk winds, jets, circumbinary disks, and outflows (e.g., \citealt{Blandford&Begelman99,Begelman+06,Blundell+08,Lu+23}). Among the most dramatic local instances of time-variable relativistic jets are the Galactic ``microquasar'' systems, most notably SS433 \citep{Fabian&Rees79,Middleton+21} and GRS 1915+105 \citep{Mirabel&Rodriguez94}. A fast precessing relativistic jet, encased in a slow shell of gas swept up by the slower disk winds, inflates egg-shaped nebulae surrounding Galactic and extragalactic ULX sources (e.g., \citealt{Pakull&Mirioni02,Roberts+03,Kaaret&Feng09}). These sources exhibit optical emission lines associated with momentum-driven shocks into the circumstellar medium (e.g., \citealt{Abolmasov+07}). Radio synchrotron emission is also produced by relativistic electrons heated or accelerated within the jet or at its termination shock on larger spatial scales of up to hundreds of pc (e.g., \citealt{Miller+05,Soria+06,Lang+07}). These local populations of X-ray binaries including ULX provide important probes of massive binary star evolution (e.g., \citealt{Kalogera+04,Pavlovskii+17}). 

However, the observed population of X-ray binaries represent a biased sample toward long-lived, stable mass transfer phases. Short lived or extreme accretion episodes, particularly those involving highly super-Eddington (``hyper-Eddington'') mass-transfer rates, are expected to be rarer and thus underrepresented in local samples. Despite their rarity in the local census, such brief phases can dominate the energy output of the system,  which power relativistic jets and inflate energetic nebulae, which -- due to their higher (``hyper-Eddington'') mass-transfer rates and more powerful jetted outflows -- are correspondingly more luminous. Such rare sources may ``outpunch their weight'' in contributing to the populations of extragalactic binary systems detectable with new sensitive wide-field surveys, particularly at radio frequencies.

Motivated by this possibility, \citet[hereafter \citetalias{Sridhar2022ApJ}]{Sridhar2022ApJ} develops a semi-analytic model for such radio ``hypernebulae'' (HNbe): nebulae inflated by extreme, short-lived mass-transfer episodes onto compact objects. Their semi-analytic framework models the dynamics of the wind-inflated expansion (e.g., \citealt{Weaver+77}) and predicts the resulting synchrotron emission from a population of electrons injected into the nebula (e.g., \citealt{Margalit&Metzger18}). \citetalias{Sridhar2022ApJ} shows that HNbe should be present in existing radio survey data, such as the Very Large Array Sky Survey (VLASS; \citealt{Lacy+20,Dong+21}). However, their detectability depends on the luminosities and lifetimes of the nebulae, which are sensitive to the properties of the underlying binary systems, particularly the mass-transfer rate, duration, and evolutionary pathway, which remain uncertain. Moreover, distinguishing HNbe from more common radio sources, such as active galactic nuclei \citep[AGN;][]{Reines+20,Eftekhari+20}, requires additional diagnostics, including host-galaxy context, spatial offsets, and potential associations with transient events. Complementary properties of these mass-transferring systems (e.g., host galaxy demographics and spatial offsets, direct detection of the donor star, or a connection to supernovae or other transient events), thus become crucial to uniquely identifying and characterizing the HNb population.

\begin{deluxetable*}{lll}
\tablecaption{Hypernebula Parameters 
(\textcolor[HTML]{66c2a5}{\textsc{Model1}}; 
\textcolor[HTML]{fc8d62}{\textsc{Model2}};
\textcolor[HTML]{8da0cb}{\textsc{Model3}})
\label{tab:modelparams}}
\tablewidth{700pt}
\tabletypesize{\scriptsize}
\tablehead{
\colhead{Symbol} & \colhead{Description} & 
\colhead{Range of Value(s)} 
} 
\startdata
$\dot{M}$ & MT rate & $10^{1} - 10^{7} \dot{M}_{\mathrm{Edd},10}$ \\
$t_{\rm acc}$ & duration of MT & $0.1-10^{6} {\rm yr}$ \\
$v_{\rm j}$ & jet velocity & $0.15-0.25c ~(\textcolor[HTML]{66c2a5}{0.15c}, \textcolor[HTML]{fc8d62}{0.2c}, \textcolor[HTML]{8da0cb}{0.25c})$ \\
$v_{\rm w}$ & disk wind velocity & $0.03c$ \\
$\varepsilon_{\mathrm{e}}$ & jet electron heating efficiency & $0.15-0.25 ~(\textcolor[HTML]{66c2a5}{0.15}, \textcolor[HTML]{fc8d62}{0.2}, \textcolor[HTML]{8da0cb}{0.25})$ \\
$\eta \equiv L_{\rm j}/L_{\rm w}$ & jet efficiency & $0.1$ \\
$\sigma_{\rm j}$ & jet/nebula magnetization & $0.1$ \\
$n_{\rm csm}$ & CSM density & $10$ cm$^{-3}$ \\
\enddata
\end{deluxetable*}

In this paper, we apply the HNb model of \citetalias{Sridhar2022ApJ} to a population of MT binaries predicted by the population synthesis code \textsc{COMPAS} \citep[version v02.35.03;][]{Riley2022ApJS}\footnote{\url{https://compas.science/}}. This allows us to consider a realistic ensemble of system properties, each with a quantifiable rate, in order to assess their detectability prospects with current or planned wide-field radio surveys, including the Very Large Array Sky Survey (VLASS; \citealt{Lacy+20}), LOFAR Two-meter Sky Survey (LoTSS; \citealt{Shimwell_2017_AA_598A_104S}), VAST \citep{Murphy2021PASA}, and DSA-2000 \citep{Hallinan2019BAAS}. 
We furthermore assess the statistical properties of the HNbe population, such as their radio luminosity function, host galaxy properties, donor star properties, and connection to explosive transients, which could aid in uniquely identifying them. Fig.~\ref{fig:schematic} summarizes the sequence of massive binary evolution leading to HNb formation.

This paper is organized as follows.  In Sec.~\ref{sec:LCmodel} we review the HNbe radio emission model. In Sec.~\ref{sec:Population} we present our binary population synthesis models and how we select a sample of hyper-Eddington accretion MT events. In Sec.~\ref{sec:Results} we present our results for the population of MT events and their resulting HNbe properties. In Sec.~\ref{sec:discussion} we discuss the observability and implications of our results for other astrophysical transients, and in Sec.~\ref{sec:conclusions} we present our conclusions.

\section{Methods}
\subsection{Hypernebula Light Curve Model}
\label{sec:LCmodel}
\subsubsection{Schematic Overview of the Model}

\begin{figure*}
\epsscale{1.17}
\plotone{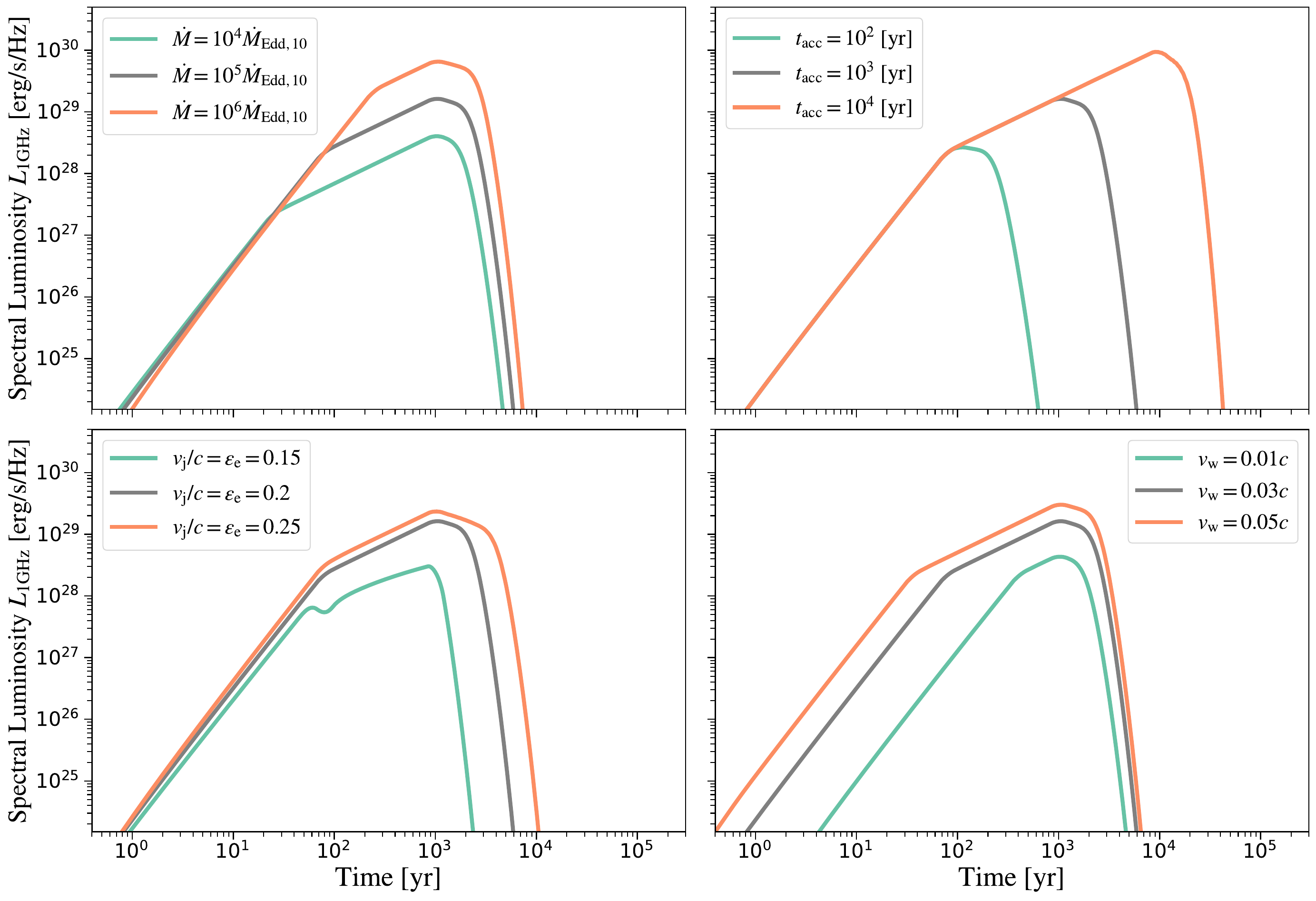}
\caption{
Example hypernebula (HNb) radio light curves (spectral luminosity at~$1\,{\mathrm{GHz}}$ as a function of time). The gray curve shows a fiducial model with parameters $(\dot{M}, t_{\mathrm{acc}}, v_{\mathrm{w}}/c, v_{\mathrm{j}}/c, \varepsilon_{\mathrm{e}}, \eta, \sigma_{\mathrm{j}},n_{\rm csm}) = (10^{5}{\dot{M}_{\mathrm{Edd},10}}, 10^{3}{\mathrm{yr}}, 0.03, 0.2, 0.2, 0.1, 0.1, 10\,{\mathrm{cm^{-3}}})$. Green and orange curves show models in which a single parameter out of eight is varied relative to the fiducial case. The panels illustrate the dependence on $\dot{M}$, $t_{\mathrm{acc}}$, $v_{\mathrm{j}}$, and $v_{\mathrm{w}}$, respectively, highlighting how these parameters primarily set the luminosity normalization and/or the duration of the radio-bright phase.
}
\label{fig:testcase_lc}
\end{figure*}

Here we briefly review the HNb model of \citetalias{Sridhar2022ApJ}.  Additional details are provided in Appendix \ref{sec:radiolc}.

The donor star of mass $M_{\rm d}$ transfers its envelope onto a compact object (CO; NS or BH) of mass $M_{\rm CO}$ at a rate $\dot{M} \gtrsim \dot{M}_{\rm Edd} \equiv L_{\rm Edd}/(0.1c^{2}),$ where $L_{\rm Edd} \simeq 1.3\times 10^{39}(M_{\rm CO}/10M_{\odot})$ erg s$^{-1}$ is the Eddington luminosity and $c$ is the light speed. The MT phase lasts a duration $t_{\rm acc} \approx \Delta M/\dot{M}$, where $\Delta M$ is the mass lost from the donor star.  The value of $t_{\rm acc}$ for a given system will depend on the state of the donor star and the binary mass ratio as reviewed in the next section.  As a result of the powerful outflows from hyper-Eddington accretion disks (e.g., \citealt{Blandford&Begelman99}), the majority of the transferred mass is ejected from the binary in a relatively slow wind of velocity $v_{\rm w} \approx 10^{4}$ km s$^{-1} \approx 0.03c$, which becomes quasi-spherical on large scales (a process aided by precession and warping of the disk, which results in self-interaction and collisions between the disk winds). This wind-fed outflow feeds a shell of radius $R_{\rm sh}$, which drives a shock into the surrounding circumstellar medium (CSM) of assumed uniform density $n_{\rm csm}$. The shell expands freely at $v_{\rm w}$ until the swept-up CSM mass becomes comparable to the injected wind mass, after which it decelerates as an energy-driven wind bubble (e.g., \citealt{Weaver+77}).

The majority of the transferred mass in hyper-Eddington systems are carried away by disk outflows. Nevertheless, a small fraction of $\dot{M}$ makes it down to small scales closest to the compact object, where it can power a (trans)relativistic ``jet'' of luminosity $L_{\rm j}$. In defining the jet, we do not distinguish between a collimated outflow directly powered by the spin energy of the compact object (e.g., \citealt{Blandford&Znajek77}) versus one powered by the innermost fastest outflows from the accretion disk \citep{Blandford&Payne82}. We assume the jet luminosity, $L_{\rm j}$, is a fraction $\eta \equiv L_{\rm j}/L_{\rm w} \lesssim 1$ of the kinetic power of the accretion disk winds, $L_{\rm w} \simeq (1/2)\dot{M}v_{\rm w}^{2}$. Since the nebula depends on the mass-transfer rate only through the absolute wind power $L_{\rm w}$ and not on $M_{\rm CO}$ explicitly, we express $\dot M$ for the radio model in units of $\dot M_{\rm Edd,10} \equiv \dot M_{\rm Edd}(10\,M_\odot)$. Values of $\eta \sim 0.1$ are motivated by the roughly equal distribution of kinetic energy in hyper-Eddington disk outflows with logarithmic radius in the disk\footnote{In particular, since the radial dependence of the mass-inflow rate scales as $\dot{M} \propto r^{p}$ in the presence of super-Eddington disk winds (e.g., \citealt{Blandford&Begelman99}, see also Appendix \ref{sec:jet_efficiency}), and the local velocity of the winds scales with the escape speed $v_{\rm w} \propto v_{\rm esc} \propto r^{-1/2}$, the kinetic power of the winds, $\propto \dot{M}v_{\rm w}^{2}/2 \propto r^{p-1}$, is weakly dependent on radius for $p \sim 1$.} and are in broad agreement with radiation hydrodynamic simulations of super-Eddington accretion (e.g., \citealt{Yoshioka+24}; see Appendix \ref{sec:jet_efficiency}). The jet expands at a velocity $v_{\rm j} \approx 0.1c-0.3c$ before terminating in a shock located near the inner edge of the swept-up wind shell (the measured jet speed in SS433 is $v_{\rm j} \approx 0.26c$; e.g., \citealt{Margon84}, see also Appendix \ref{sec:ss433}).

Such magnetized (trans-)relativistic shocks are assumed to place a fraction $\varepsilon_{\mathrm{e}} \sim 0.1-0.5$ of the dissipated jet power into heating electrons with a relativistic Maxwellian distribution (e.g., \citealt{Sironi&Spitkovsky11}).\footnote{Ions may also be heated and/or accelerated into a non-thermal power-law energy distribution at the termination shock. This makes HNbe sources of cosmic rays and high-energy neutrinos via photohadronic interactions \citep{Sridhar+24a}.}  The injection of relativistic particles inflates a magnetized ``nebula'' behind the wind termination shock whose radius $R_{\rm n} < R_{\rm sh}$ is defined by a balance between the ram pressure of the jet and the pressure within the nebula. The magnetic field injected into the nebula is set by the magnetization of the jet, $\sigma_{\rm j}$.  The energy distribution of the electrons in the nebula evolves in time due to various expansion and radiation losses. The latter includes radio synchrotron emission, which is observable for a duration $\simeq t_{\rm acc}$ commensurate with the MT timescale over which the nebula remains energized by the jet.

\subsubsection{Parameter Dependencies and Model Grid}

\begin{figure*}
\epsscale{1.17}
\plotone{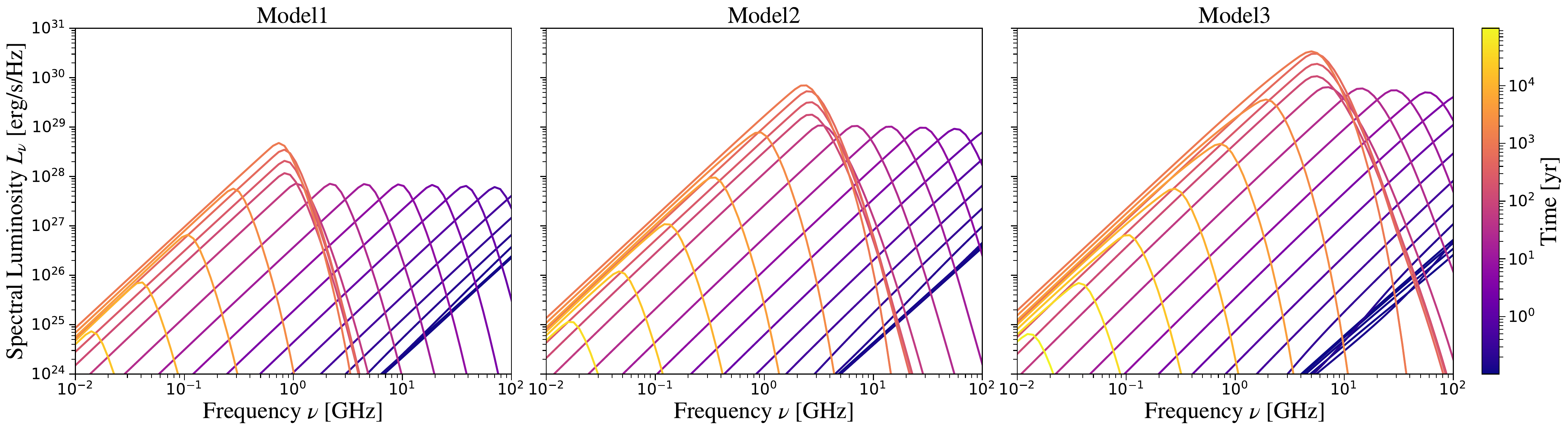}
\caption{Snapshots of the spectral energy distributions for each of the three HNb model grids summarized in the rightmost column of Table \ref{tab:modelparams}, color-coded by the system age. The key model parameters are $(\dot{M}/\dot{M}_{{\rm Edd},10}, t_{\mathrm{acc}}/{\rm yr}, v_{\rm j}/c, \varepsilon_{\mathrm{e}}) = (10^{5}, 10^{3}, 0.15, 0.15)$ for \textsc{Model1}, $(10^{5}, 10^{3}, 0.2, 0.2)$ for \textsc{Model2}, and $(10^{5}, 10^{3}, 0.25, 0.25)$ for \textsc{Model3}.} 
\label{fig:sed}
\end{figure*}

\begin{figure*}
\epsscale{1.17}
\plotone{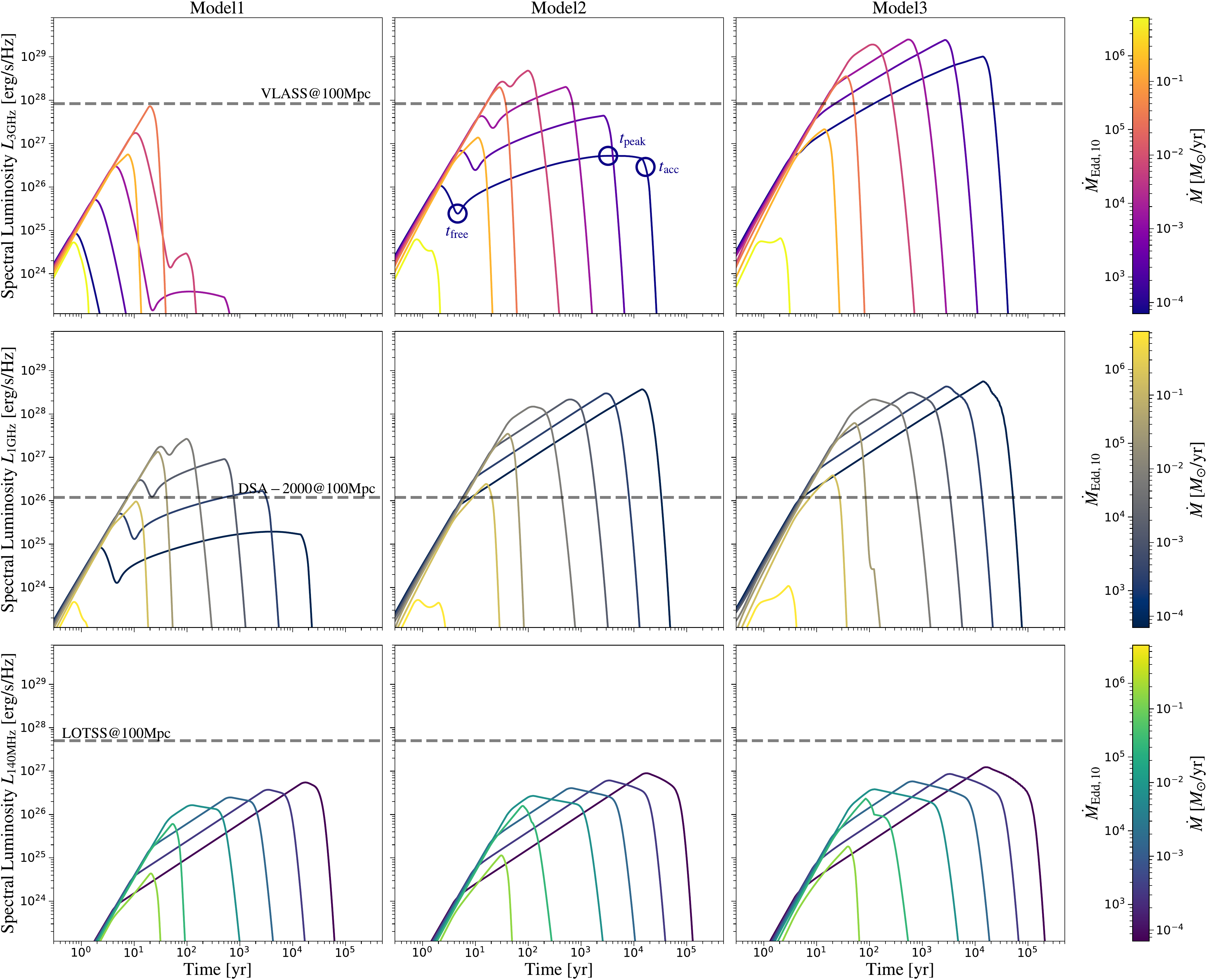}
\caption{Light curves for each of the three models summarized in the rightmost column of Table \ref{tab:modelparams}, color-coded by $\dot{M}$ (in unit of $M_{\odot}$ yr$^{-1}$ and $\dot{M}_{\mathrm{Edd},10}$). The accretion timescale $t_{\rm acc}$ follows the mean relationship derived from our sample of hyper-Eddington mass-transfer events in absolute $\dot M$; $\log(t_{\rm acc}/{\rm yr}) = -0.27 - 1.08\log(\dot M/M_\odot\,{\rm yr^{-1}})$. The top, middle, and bottom panels show $3$ GHz, $1$ GHz, and $140$ MHz light curves corresponding to VLASS, DSA--2000, and LoTSS, respectively (see Table \ref{tab:survey}). In the top-center panel, the typical timescales of $t_{\mathrm{free}}$, $t_{\mathrm{peak}}$, and $t_{\mathrm{acc}}$ are highlighted.}
\label{fig:lc_comp}
\end{figure*}

As summarized in Table \ref{tab:modelparams}, a given HNb model is characterized by 8 parameters: $\{\dot{M}, t_{\mathrm{acc}}, v_{\rm w}, v_{\rm j}, \varepsilon_{\mathrm{e}}, \eta, \sigma_{\rm j},n_{\rm csm}\}$. Fig.~\ref{fig:testcase_lc} shows a series of $1 {~\mathrm{GHz}}$ radio light curves spanning a range of these parameters, with $(10^{5}\dot{M}_{\mathrm{Edd},10}, 10^{3}{\,\mathrm{yr}}, 0.03c, 0.2c, 0.2, 0.1, 0.1, 10)$ adopted as the fiducial parameter set. A typical light curve rises for decades to centuries, increasing until a timescale of order $t_{\mathrm{acc}}$, after which the jet shuts off and the luminosity drops sharply. The peak luminosity generally increases for higher MT rates $\dot{M}$, higher jet velocities $v_{\rm j}$, and higher wind velocities $v_{\rm w}$.

We calculate three grids of models, each sampling a dense grid of $\{\dot{M}, t_{\mathrm{acc}}\}$, as summarized in the right-hand column of Table \ref{tab:modelparams}. Each MT episode in our population (Sec.~\ref{sec:Population}) is then assigned the light curve computed at the center of its cell in this logarithmic grid. Motivated by the jet velocity sensitivity for the light curves (Fig.~\ref{fig:testcase_lc}), the three models are distinguished by the adopted jet velocity and electron heating efficiency: \textsc{Model1}: ``low-$v_{\rm j}$'' ($v_{\rm j}/c = \varepsilon_{\mathrm{e}}=0.15$); \textsc{Model2}: ``moderate-$v_{\rm j}$'' ($v_{\rm j}/c = \varepsilon_{\mathrm{e}}=0.2$); \textsc{Model3}:``high-$v_{\rm j}$'' ($v_{\rm j}/c = \varepsilon_{\mathrm{e}}=0.25$).

We vary $v_{\rm j}$ instead of $v_{\rm w}$ because of its larger uncertainty (the jet speed depends on poorly known conditions close to the accretor, such as the magnetic field strength threading the black hole or inner accretion disk), and because $v_{\rm j}$ is further coupled to the uncertain parameter $\varepsilon_{\mathrm{e}}$ (see Eq.~\eqref{eq:Ngamma}). We also increase $\varepsilon_{\mathrm{e}}$ in tandem with $v_{\rm j}$ across the models, motivated by kinetic simulations of trans-relativistic collisionless shocks which show higher electron heating at trans-relativistic shocks with greater speeds (e.g., \citealt{SironiTran24}). The other parameters are fixed at characteristic values: $v_{\rm w} = 0.03c$, $\eta=0.1$, $\sigma_{\rm j} = 0.1$, $n_{\mathrm{csm}} = 10$ cm$^{-3}$, independent of $\dot m$ and $M_{\rm CO}$ (see \citetalias{Sridhar2022ApJ}).

Fig.~\ref{fig:sed} shows the nebula spectral energy distribution (SED) at different snapshots in time, for each of three model grids spanning a range of $\dot{M}/\dot{M}_{{\rm Edd},10}$ and $t_{\rm acc}$. The evolution of the SED is controlled by two physical timescales: (1) $t_{\rm free}$, when the nebula transitions from free expansion to energy-driven (decelerating) evolution, and (2) $t_{\rm acc}$, the duration of the MT phase (for details, see \citetalias{Sridhar2022ApJ}). As the nebula expands, its magnetic field weakens and the peak frequency of the synchrotron SED shifts to lower frequencies until $t=t_{\mathrm{free}}$, while the total power remains approximately constant. Finally, after the fresh injection of electrons from the jet terminates at $t = t_{\mathrm{acc}}$, both the total power of the SED and the peak frequency rapidly decay. For the light curve at a given observing frequency, the interplay of $t_{\rm free}$ and $t_{\rm acc}$ shapes its overall evolution: the emission initially rises (and declines, depending on the observing frequency) until $t=t_{\mathrm{free}}$, followed by a more slowly evolving phase until $t=t_{\mathrm{acc}}$, after which the light curve rapidly fades. In addition, we further define a third characteristic timescale, $t_{\rm peak}$, as the time at which the light curve reaches its maximum at a given observing frequency (see also Fig.~\ref{fig:lc_comp}).

The non-monotonic dependence of the observed light curves on mass-transfer rate in Fig.~\ref{fig:lc_comp} reflects the interplay between increasing luminosity and a shifting SED peak. As the mass-transfer rate increases, the overall luminosity rises, but the accompanying shift in the SED peak can counteract this brightening in a given observing band. The low jet speed and electron heating efficiency of \textsc{Model1}, results in the SED peaking at low frequencies $\sim 800 {\mathrm{~MHz}}$, close to the bandpass of the VLASS and LoTSS surveys (see Fig. \ref{fig:sed} left). For \textsc{Model1}, higher frequency observers probe the optically-thin portion of the SED, while those at lower frequencies see the optically-thick self-absorbed tail, thus leading to the distinct behavior of the 140 MHz and 3 GHz light curves for \textsc{Model1}. By contrast, the higher $v_{\rm j}$ and $\varepsilon_{\mathrm{e}}$ of \textsc{Model3} (see Fig. \ref{fig:sed} right) results in a higher SED peak at $\sim 5$ GHz, such that both low- and high-frequency observations will see the optically-thick part of the SED, resulting in a similar light curve shape.

\subsection{Binary Population Synthesis} \label{sec:Population}

\subsubsection{Overview of \textsc{COMPAS} Simulation}
\label{subsec:compas}

To simulate a population of binary systems undergoing hyper-Eddington MT phases, we employ the rapid population synthesis code \textsc{COMPAS} \citep{Riley2022ApJS}. In particular, we use the \textsc{COMPAS} dataset calculated by \citet{vanSon+22} of $10^{7}$ binaries with zero age main sequence (ZAMS) primary star masses in the range $5-150 M_{\odot}$. The primary stars follow the Kroupa initial mass function \citep[IMF,][see also Fig. \ref{fig:mass_distribution_zams}]{Kroupa01}, while the secondaries follow a uniform mass distribution with respect to the primary.

An adaptive importance sampling is employed using STROOPWAFEL \citep{Broekgaarden2019MNRAS}, which improves the computational efficiency of population studies of rare events, by focusing the simulation around regions of the initial parameter space found to produce outputs of interest. \citet{vanSon+22} was focused on studying binary black hole formation, so they densely sample the stellar mass ranges around $\sim 30 M_{\odot}$ which predominately birth black holes; however, we correct for this sampling bias in the following results.

We sample birth metallicities from a logarithmically flat distribution in the range of $10^{-4} \leq Z \leq 0.03$. We account for the metallicity-dependence of the star formation history in calculating binary formation rates by re-weighting the simulated population, as described below in Sec.~\ref{sec:rates} \citep[for more details, see][]{vanSon+22, Broekgaarden2019MNRAS}.

\subsubsection{Stable and Unstable Mass Transfers} \label{subsubsec:mt}

The compact object accretion events giving rise to HNbe can arise from either stable MT (i.e., Roche lobe overflow; RLOF) or unstable MT (i.e., common envelope; CE), as illustrated by stage 5 in Fig.~\ref{fig:schematic}. In \textsc{COMPAS}, MT is assumed to occur for a given binary when the condition $R_{\mathrm{d}} = R_{\mathrm{RL}}(1-e)$ is satisfied, where $R_{\mathrm{d}}$ and $R_{\mathrm{RL}}$ are the radius and Roche radius of the donor, respectively, and $e$ is the binary eccentricity. The Roche radius is estimated as \citep{Eggleton1983ApJ}:
\begin{equation}
    R_{\mathrm{RL}} = 0.49 \frac{q_{\mathrm{RL}}^{2/3}}{0.6 q_{\mathrm{RL}}^{2/3} + \ln(1+q_{\mathrm{RL}}^{1/3})} a,
\end{equation}
where $q_{\mathrm{RL}} = M_{\mathrm{d}} / M_{\mathrm{CO}}$ is the ratio of the donor mass ($M_{\mathrm{d}}$) to the compact object mass ($M_{\mathrm{CO}}$), and $a$ is the binary separation.

Whether a given MT event is classified as stable or unstable depends on the response of the donor star radius $R_{\mathrm{d}}$ to mass-loss relative to its Roche lobe size, e.g., by the parameters $\zeta_{\mathrm{RL}} \equiv d\ln (R_{\mathrm{RL}}) / d\ln (M_{\mathrm{d}})$ and $\zeta_{\mathrm{d}} \equiv d\ln (R_{\mathrm{d}}) / d\ln (M_{\mathrm{d}})$, following commonly used analytic approximations \citep{Soberman1997AA, Riley2022ApJS}.  As long as $\zeta_{\mathrm{RL}} \leq \zeta_{\mathrm{d}}$, the MT is stable/RLOF, while unstable/CE systems occur for $\zeta_{\mathrm{RL}} \geq \zeta_{\mathrm{d}}$. Because the donor loses mass during MT ($dM_{\mathrm{d}} < 0$), this criterion physically corresponds to the donor radius contracting more rapidly ($\zeta>0$) -- or expanding more slowly ($\zeta<0$) -- than its Roche lobe in response to mass loss. In the stable case, the donor radius retreats within its Roche lobe, while in the unstable case, it overflows the Roche lobe.

In stable RLOF cases, the MT rate in \textsc{COMPAS} is determined by the thermal timescale of the donor:
\begin{equation}
    \dot{M}_{\mathrm{th}} = \frac{M_{\mathrm{d, env}}}{\tau_{\mathrm{KH}}} = \frac{R_{\mathrm{d}} L_{\mathrm{d}}}{G M_{\mathrm{d}}}, \qquad
    \tau_{\mathrm{KH}} = \frac{G M_{\mathrm{d}} M_{\mathrm{d, env}}}{R_{\mathrm{d}} L_{\mathrm{d}}},
\end{equation}
where $L_{\mathrm{d}}$ is the donor luminosity, and $\tau_{\mathrm{KH}}$ is the Kelvin-Helmholtz (thermal) timescale of the donor envelope. \textsc{COMPAS} limits the mass growth rate of accreting compact objects to the Eddington rate. Assuming that most of the accreting mass is ejected as the disk wind due to the hyper-Eddington accretion, we take the MT rate to be the mass-loss rate of the donor:
\begin{equation}
    \dot{M} = \dot{M}_{\mathrm{th}}, \quad
    t_{\mathrm{acc}} = \frac{\Delta M_{\mathrm{RLOF}}}{\dot{M}_{\mathrm{th}}} = \tau_{\mathrm{KH}},\quad\text{RLOF},
\end{equation}
where $\Delta M_{\mathrm{RLOF}} \equiv M_{\mathrm{d, preRLOF}} - M_{\mathrm{d, postRLOF}}$ is the change in the donor mass between the pre- and post-RLOF phases, which here equals the envelope mass $M_{\mathrm{d, env}}$, and $t_{\mathrm{acc}}$ is the duration of the accretion phase.

For donors without a core--envelope structure (main-sequence and helium main-sequence stars), \textsc{COMPAS} resolves the MT over many time steps, keeping the donor within its Roche lobe. In this case, we use the resolved rate, capped at the thermal rate:
\begin{equation}
    \dot{M} = \min\left(\frac{\Delta M_{\mathrm{RLOF}}}{\Delta t},\, \dot{M}_{\mathrm{th}}\right),\,\,\,\text{RLOF (no core)},
\end{equation}
where $\Delta t$ is the duration of the \textsc{COMPAS} time step. Consecutive time steps of the same binary that all exceed our hyper-Eddington threshold are combined into a single accretion episode, with $\dot{M} = \sum \Delta M_{\mathrm{RLOF}} / \sum t_{\mathrm{acc}}$.

\textsc{COMPAS} treats unstable CE phases following the standard $\alpha_{\mathrm{CE}}-\lambda$ formalism \citep{Webbink1984ApJ, deKool1990ApJ}, in which the initial gravitational binding energy of the donor envelope is related to the change in the binary orbital energy according to $E_{\mathrm{bind}} = \alpha_{\mathrm{CE}} \Delta E_{\mathrm{orb}}$, where $\alpha_{\mathrm{CE}}$ is the CE parameter.  The envelope binding energy is expressed by $E_{\mathrm{bind}} = - GMM_{\mathrm{env}}/\lambda R$, where $\lambda$ is a structure parameter \citep{Webbink1984ApJ}. Following the default setup in \textsc{COMPAS}, we adopt $\alpha_{\mathrm{CE}} = 1.0$ and `Nanjing' value for $\lambda$ \citep{Xu2010ApJ}. The energy balance of the CE systems involving compact object accretors is thus determined as follows:
\begin{equation}
- \frac{G M_{\mathrm{d}} M_{\mathrm{d, env}}}{\lambda_{\mathrm{d}} R_{\mathrm{d}}} =\alpha_{\mathrm{CE}}\left(\frac{G M_{\mathrm{d}} M_{\mathrm{CO}}}{2 a_{\mathrm{preCE}}}-\frac{G M_{\mathrm{d, core}} M_{\mathrm{CO}}}{2 a_{\mathrm{postCE}}}\right),
\end{equation}
where $a_{\mathrm{pre/postCE}}$ are the binary separations of the pre-/post-CE phases, and the subscripts `env' and `core' refer to the envelope and core mass, respectively. The criterion for successful envelope ejection is $a_{\mathrm{postCE}} > R_{\mathrm{d}} + R_{\mathrm{CO}}$, where $R_{\mathrm{CO}}$ is the CO radius (taken to be $10 {\mathrm{~km}}$ for neutron stars and the Schwarzschild radius for black holes). We further adopt the ``pessimistic'' CE assumption: Hertzsprung-gap (HG) donors, which lack a well-developed core--envelope boundary, are assumed not to survive a CE. Likewise, binaries in which the stripped donor immediately overfills its Roche lobe after the CE are assumed to merge. We also count as mergers those binaries in which an HG donor expands beyond the binary separation within a single evolutionary time step of stable MT, which \textsc{COMPAS} records as a stellar merger without a CE. We exclude all merging systems from our sample (see also Fig.~\ref{fig:sankey}).

\textsc{COMPAS} assumes that any unstable CE phase occurs instantaneously, insofar that the runaway MT and its resulting plunge into, and ejection of, the envelope (e.g., \citealt{MacLeod+18,Lau+22}) take place much faster than other stages of stellar evolution. Once the accretor fully enters and begins to spiral in through the donor's envelope, it is not clear how much additional mass it will gain (e.g., \citealt{MacLeod+17}), nor whether an accretion-powered jet from the compact object will be able to escape from the envelope (e.g., \citealt{Hillel+23}). Nevertheless, significant disk-fed CO accretion and jet-production will occur {\it prior to} the CO entering the donor envelope (e.g., \citealt{Soker15}).  This occurs for at least tens or hundreds of orbits as the MT rate rises during the runaway process. Such a ``pre-dynamical'' MT phase is supported observationally in stellar mergers by the gradually-rising pre-maximum light curves of luminous red novae (e.g., \citealt{Soker&Tylenda06,Tylenda+11,Pejcha+17,Matsumoto&Metzger22}).

Again motivated by stellar merger observations \citep{Pejcha+17}, we assume that a fraction $f_{\rm acc} \sim 0.1$ of the donor envelope mass removed during the CE is channeled towards the CO during this pre-dynamical phase, over the course of $100$ orbital periods, $P_{\mathrm{orb}}$, which are calculated by $2\pi \sqrt{a^3 / (G (M_{\mathrm{d}}+M_{\mathrm{CO}}))}$. The MT rate for CE phases is thus determined from the \textsc{COMPAS} output as follows:
\begin{equation}
    \dot{M} = \frac{f_{\mathrm{acc}} \Delta M_{\mathrm{CE}}}{100 \times P_{\mathrm{orb}}},\,\,\,\text{CE}
\end{equation}
where $\Delta M_{\mathrm{CE}}$ is the change in donor star mass during the CE. Although this prescription is highly uncertain, CE systems contribute only $\lesssim 2\%$ of the observable HNb population in our adopted models.

\subsubsection{Defining the Hyper-Eddington Sample} \label{subsubsec:sampling}

From our sample of $10^{7}$ evolved binary systems, we first separate all of the MT events into RLOF versus CE. From these, we then define our sample of ``hyper-Eddington'' events as those above a critical Eddington ratio:
\begin{equation}
    \label{eq:def_hnb}
	\dot{m} \equiv \frac{\dot{M}_{\mathrm{RLOF/CE}}}{\dot{M}_{\rm Edd}(M_{\rm CO})} \geq 100.
\end{equation} 
Note that the selection criterion uses the Eddington ratio of each system, since the physics of the accretion flow is governed by $\dot m$, whereas the radio light curves depend on the absolute $\dot M$ (Sec.~\ref{sec:LCmodel}).

\begin{figure*}
\centering
\epsscale{1.17}
\plotone{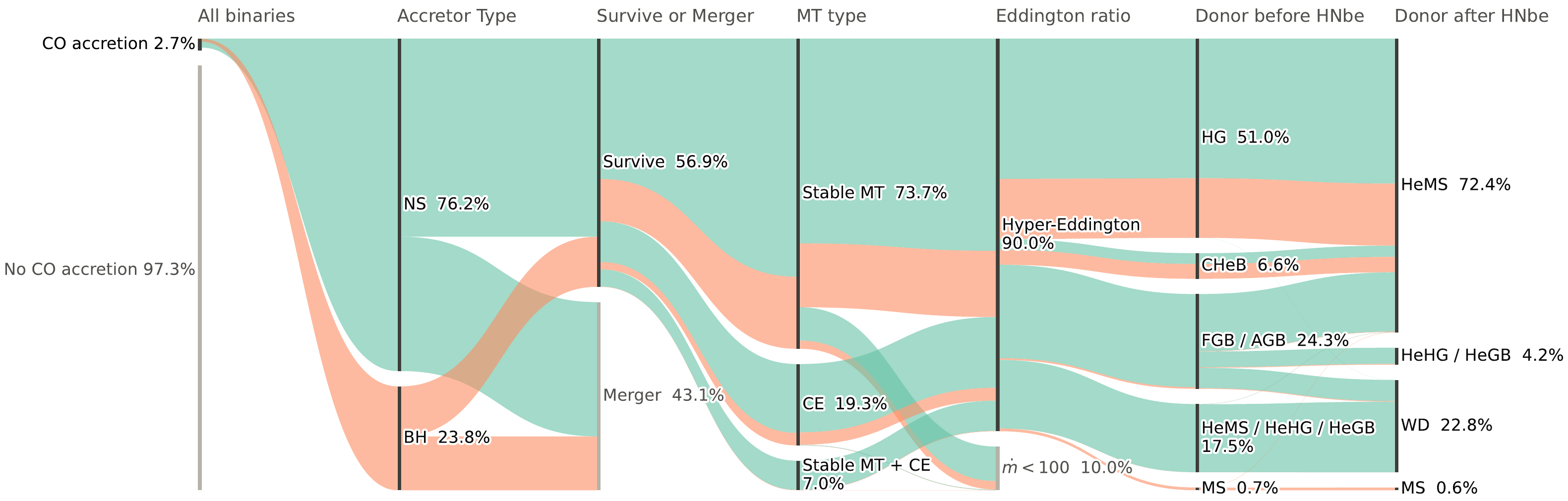}
\caption{
Paths to the HNb population at each selection criterion in our COMPAS population, weighted by the formation rate in the local universe ($z \approx 0$). Band colors show the accretor type (green: NS; orange: BH). Grey nodes mark systems removed at that step. The last two columns show the donor type before and after the first hyper-Eddington episode of each system. Because a system can undergo multiple hyper-Eddington episodes, these fractions differ from the episode-based fractions in Fig.~\ref{fig:beforeaftermapping}. Stellar types follow \citet[][see Sec. \ref{subsec:population}]{Hurley2000MNRAS}. 
}
\label{fig:sankey}
\end{figure*}

Fig.~\ref{fig:sankey} traces how the hyper-Eddington sample is extracted from the full \textsc{COMPAS} population, weighting each binary by its formation rate in the local universe ($z \approx 0$, see Sec. \ref{sec:rates}). Only 2.7\% of all binaries undergo mass transfer onto compact objects. Under the pessimistic CE assumption, 43\% of these systems end in a merger. For the surviving systems, 90\% of them reach the Hyper-Eddington phase in at least one episode.

\begin{figure}
\centering
\epsscale{1.17}
\plotone{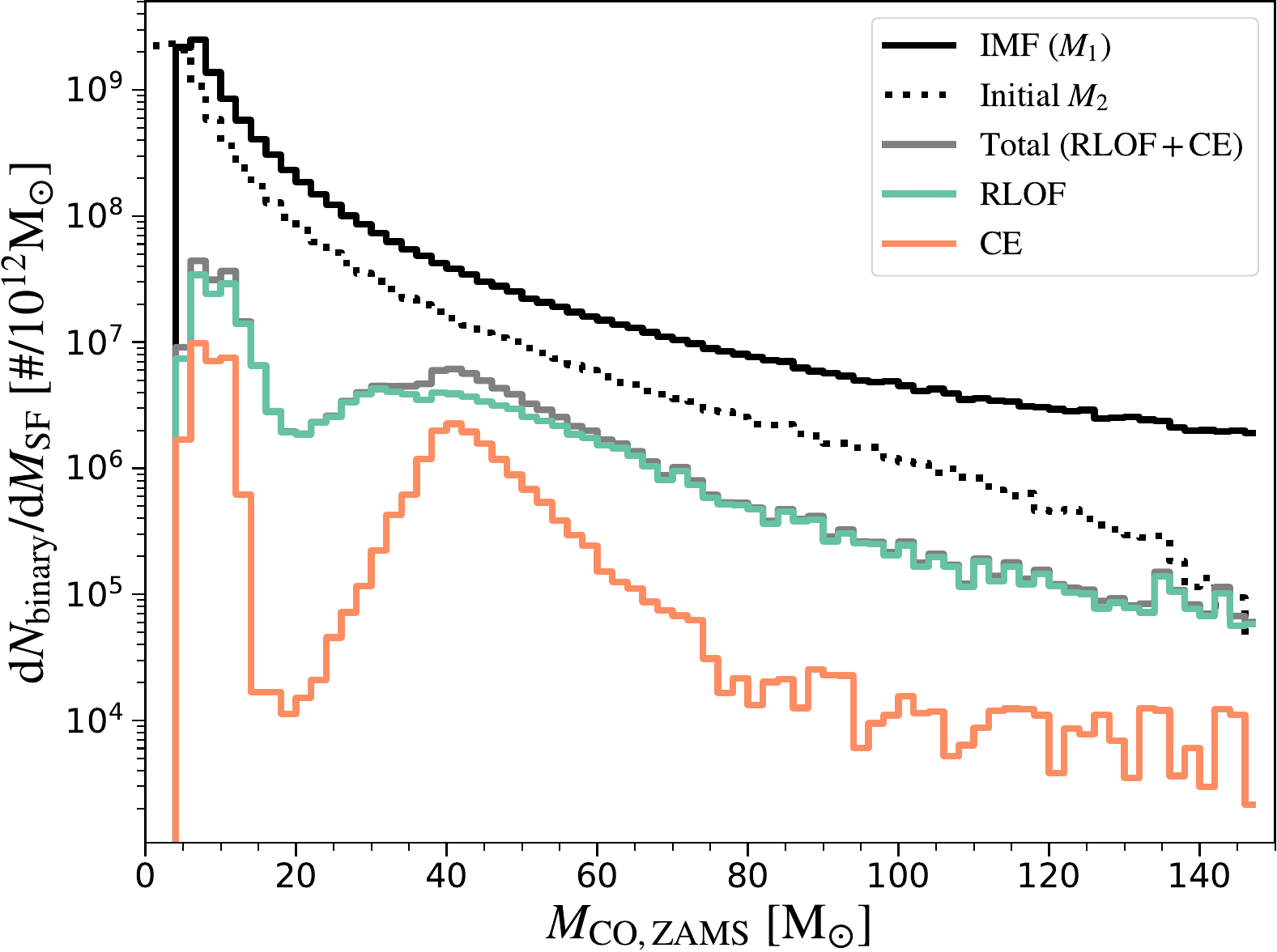}
\caption{Distribution of ZAMS mass of the CO accretor for the initial stellar distribution for the COMPAS simulation (IMF), the secondary star distribution, and of each MT type as a function of the ZAMS mass specifically leading to the hyper-Eddington MT systems. Black solid and dotted lines show the initial population of binaries, while gray line shows the total population at the onset of MT phases. The latter is also broken down into systems undergoing stable RLOF (green curves) versus unstable CE (orange curves).}
\label{fig:mass_distribution_zams}
\end{figure}

Fig.~\ref{fig:mass_distribution_zams} shows the ZAMS mass distribution of the progenitors of the accreting COs in our hyper-Eddington sample, normalized to the star-forming mass of $10^{12}\,M_{\odot}$ (assuming a binary fraction of 0.7) and combined over all metallicities. The accretor is the initially more massive (primary) star in all CE episodes and in most ($\sim 85\%$) of RLOF episodes. In the remaining RLOF episodes, mass-ratio reversal during an earlier MT phase allows the initially less massive star to collapse first and become the accretor. While the IMF simply follows the Kroupa mass function \citep[$\propto M^{-2.3}$;][]{Kroupa01}, the hyper-Eddington systems exhibit a more complex distribution. The first peak at $\sim 10\,M_{\odot}$ corresponds to accretion onto NSs, and the second peak at $\sim 40\,M_{\odot}$ to accretion onto BHs. The dip between them is shaped by the natal kick at the first SN. Below $\sim 12\,M_{\odot}$, most primaries also explode as core-collapse SNe, and the binaries are disrupted. However, $\sim 10\%$ of them explode as electron-capture or ultra-stripped SNe, for which \textsc{COMPAS} assumes weak kicks. Because $50$--$80\%$ of these binaries remain bound, they account for most of the NS-accreting systems in our sample. Instead, the progenitors of $\sim 15$--$30\,M_{\odot}$ undergo core-collapse SNe with little fallback and receive nearly the full kick. Only a few percent of these binaries remain bound, mostly tight post-CE systems in which the massive companion soon fills its Roche lobe and, in most cases, merges with the CO. Above $\sim 30\,M_{\odot}$, nearly complete fallback suppresses the kick, so that most binaries remain bound and can later undergo hyper-Eddington MT.

\subsubsection{Rates Across Cosmic Time}
\label{sec:rates}

We assume that the time delay from the birth of a massive star binary to a hyper-Eddington MT phase is negligible compared to cosmic evolution timescales. The birth rate of HNbe at a given redshift is therefore taken to be proportional to the star formation rate density (SFRD) according to:
\begin{equation}
    \mathcal{R} (z) = \int_{Z_{\min}}^{Z_{\max}} \mathcal{S} (Z^{\prime}, z) \times \frac{dN_{\mathrm{binary}}}{dM_{\mathrm{SF}}}dZ^{\prime},
\end{equation}
where $dN_{\mathrm{binary}}/dM_{\mathrm{SF}}$ is the number of binaries per unit star-forming mass.
Here, 
\begin{align}
    \mathcal{S} (Z, z) &\equiv \frac{d^3 M_{\mathrm{SF}}}{dt dV dZ} (Z, z) \\
    &= \frac{d^2 M_{\mathrm{SF}}}{dt dV} (z) \times \frac{d P_{\mathrm{SF}}}{d Z} (Z, z), 
    \label{eq:S}
\end{align}
is the metallicity-specific SFRD. The first term entering Eq.~\eqref{eq:S} is the star formation rate density, i.e., the amount of star-forming gas per unit time per unit volume \citep{Madau2014ARAA}, while the second multiplicative term is the metallicity distribution function, i.e., the probability of star formation at metallicity, $Z$, and redshift, $z$ \citep[e.g.,][]{Neijssel2019MNRAS, vanSon+23}. Additional details are provided in Appendix \ref{sec:sfrd}.

\section{Results}
\label{sec:Results}

Before presenting the quantitative results, we note that, although our framework enables explicit predictions for the MT-event population, HNb properties, and detectable source counts, these estimates remain sensitive to several uncertain modeling assumptions. Our calculation combines a simplified semi-analytic radio emission model, an adopted mapping between accretion and outflow properties, and assumptions entering the binary population synthesis calculation, particularly for common-envelope systems and their pre-dynamical accretion phase. Because these ingredients enter nonlinearly, modest changes in parameters such as the jet velocity or electron heating efficiency can shift the predicted detectable number by orders of magnitude. Accordingly, in this section and the discussion that follows, we focus primarily on robust qualitative trends and the overall scale of the expected population.

We now describe the population of hyper-Eddington MT systems (Sec.~\ref{subsec:population}), before characterizing their resulting HNb properties (Sec.~\ref{subsec:HNbpopulation}) and their detectability in radio surveys (Sec.~\ref{subsec:surveydetect}).

\subsection{Population of MT Events}
\label{subsec:population}

\begin{figure}
\centering
\epsscale{1.17}
\plotone{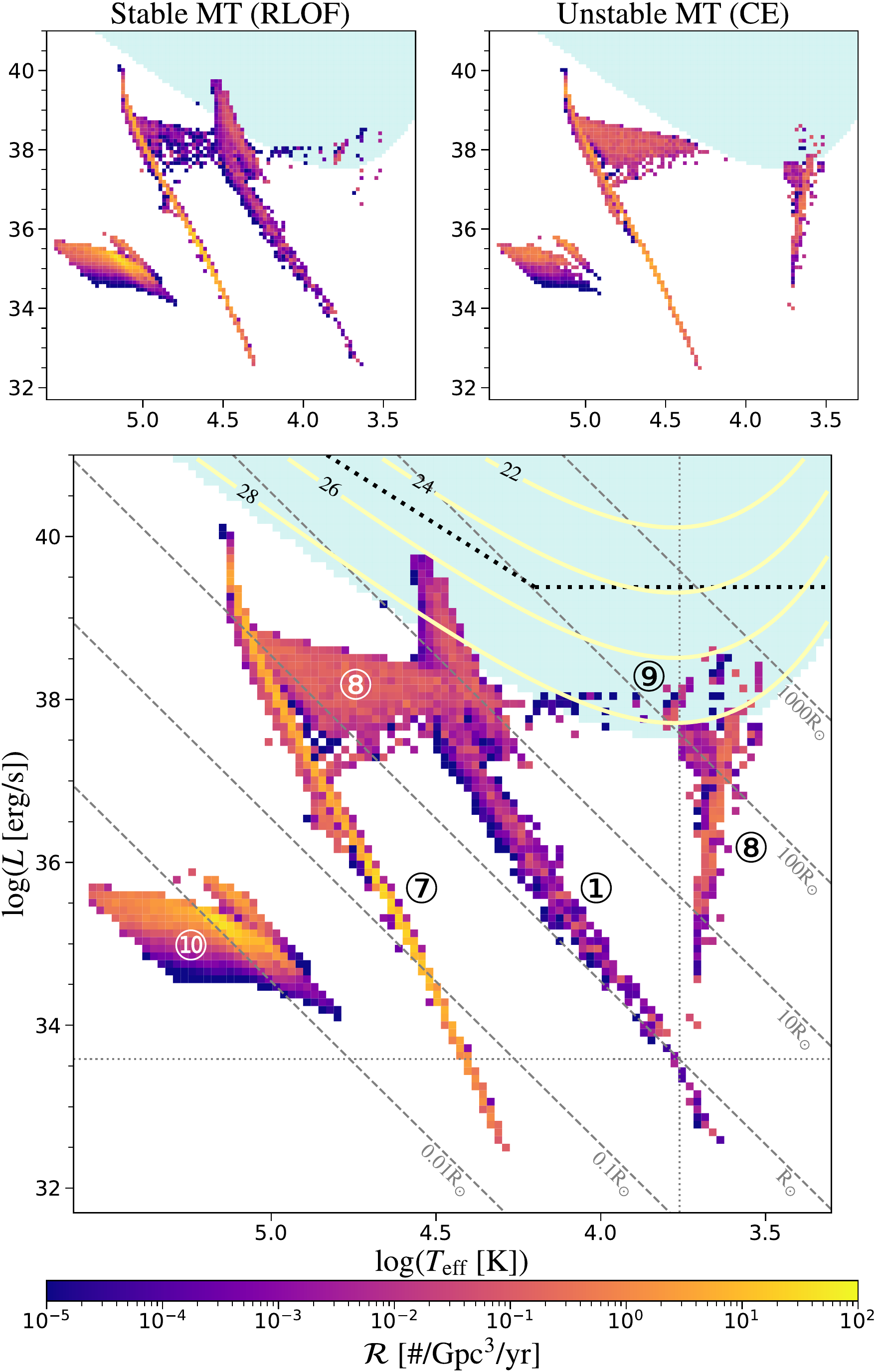}
\caption{Hertzsprung–Russell (HR) diagrams of donor stars following hyper-Eddington MT phases $\dot{m} \geq 10^{2}$, color-coded by the volumetric event rate. The main lower panel shows the entire population of the donor stars in binaries, which is further broken down in the top sub-panels into stable RLOF versus CE MT phases. Circled numbers on the plot label the phase of the donor stellar evolution, as outlined in the main text. The gray dotted lines show the solar luminosity and temperature, while diagonal dashed lines show the blackbody radii. The blue-shaded region denotes donor stars detectable by the {\it Nancy Grace Roman Space Telescope}, assuming a blackbody spectrum, source distance of $30{~\mathrm{Mpc}}$, and the F062 filter with a 1-hour exposure (yellow contours denote apparent magnitude in the F062 filter). The black dotted line is the Humphreys-Davidson limit \citep{Humphreys1979ApJ_232_409H}.
}
\label{fig:hr}
\end{figure}

\begin{figure*}
\epsscale{1.17}
\plotone{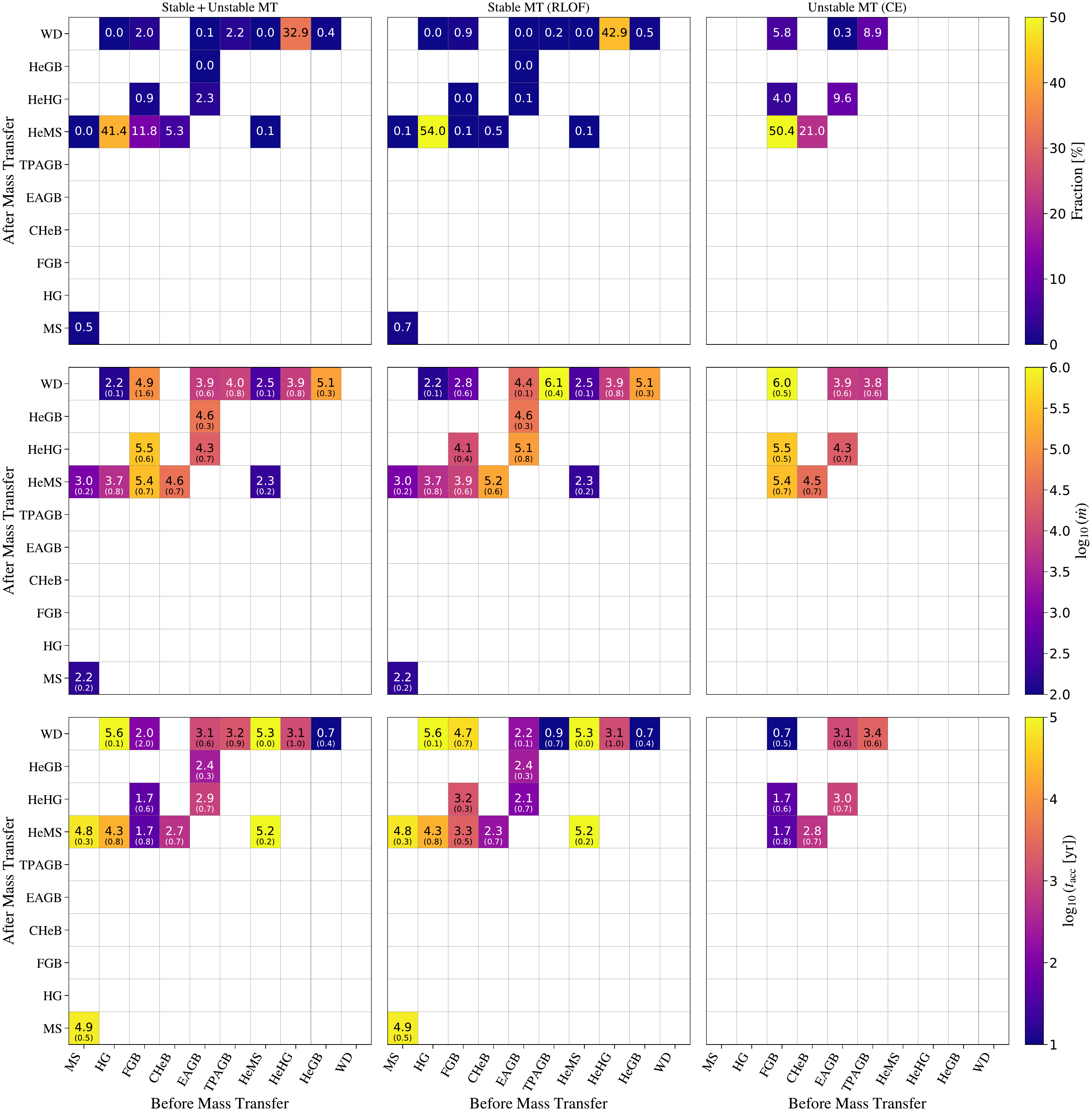}
\caption{Mapping between the state of the donor star before and after MT phases, showing the fraction of all hyper-Eddington sources (top panels), mean (standard deviation) of the accretion rate $\dot{m}$ (middle panels), and MT duration (bottom panels). The left column shows the all of hyper-Eddington sources, and the center and right columns break the sample into stable RLOF and unstable CE events, respectively.} 
\label{fig:beforeaftermapping}
\end{figure*}

Fig.~\ref{fig:hr} shows the Hertzsprung-Russell (HR) diagram of the donor stars after the MT event, colored according to volumetric rate in the local universe ($z \approx 0$). The panels above the main figure separate these systems into those that underwent RLOF versus CE. Based on the categories proposed by \citet{Hurley2000MNRAS}, the donor stars are divided into the following evolutionary stages:
\begin{enumerate}
    \item Hydrogen Main-Sequence (MS),
    \item Hertzsprung Gap (HG),
    \item First Giant Branch (FGB),
    \item Core Helium Burning (CHeB),
    \item Early AGB (EAGB),
    \item Thermally Pulsational AGB (TPAGB)
    \item Helium Main-Sequence (He),
    \item Helium Hertzsprung Gap (HeHG),
    \item Helium Giant Branch (HeGB),
    \item White Dwarf (WD).
\end{enumerate}

Because binaries that undergo MT leading to stellar types $2-6$ are assumed to merge and hence do not survive as binary systems, they are not labeled in Fig.~\ref{fig:hr}. For example, Case A MT generally leads to stable RLOF with relatively slow transfer rates driven by the long thermal timescale of MS stars. On the other hand, although a fraction of Case B MT systems survive through stable MT (e.g., \citealt{Pavlovskii+17}), they frequently become unstable due to the deep convective envelope of the HG donor.

Luminous HeHG and HeGB donors occupy the upper-right corner of the HR diagram. Figure~\ref{fig:hr} also highlights donors detectable as optical/near-infrared counterparts to HNbe (see also Sec.~\ref{sec:discussion} and \citealt{Blanchard_2025_ApJ_989_L49}). The blue shaded region indicates donors bright enough to be detected at a distance of 30 Mpc in a 1-hour exposure with the F062 filter on the {\it Nancy Grace Roman Space Telescope} ({\it Roman}). For comparison, \citet{Blanchard_2025_ApJ_989_L49} assess donor detectability at 40 Mpc using the F150W filter on the {\it James Webb Space Telescope} ({\it JWST}), for which a 2.7-hour exposure reaches a $5\sigma$ limiting magnitude of $\sim 30$.

Figure~\ref{fig:beforeaftermapping} maps the evolutionary states of donors before and after MT. For each before$\rightarrow$after channel, we report its fractional contribution to the total population of hyper-Eddington systems, together with its mean MT rate $\dot{m}$ and duration $t_{\rm acc}$. The dominant channels are HG$\rightarrow$HeMS and HeHG$\rightarrow$WD, which account for $\sim 41\%$ and $\sim 33\%$ of all systems, respectively. We also separate the population into RLOF and CE events. As expected, CE events typically have much shorter durations and higher MT rates than RLOF events.

Figure~\ref{fig:rate_2d} shows the volumetric birth rate of MT events in the $\{\dot{m},\,t_{\rm acc}\}$ plane. Consistent with Fig.~\ref{fig:beforeaftermapping}, shorter MT episodes typically have higher MT rates. The population is broadly concentrated around the scaling relation
\begin{equation}
\dot{m}^{1.25\pm0.05} \sim 10^{5.33 \pm 0.23}\left(\frac{t_{\rm acc}}{10^{3}{\rm yr}}\right)^{-1}.
\label{eq:Mdot_tacc_scaling}
\end{equation}
We obtain this relation by fitting the non-zero bins in the two-dimensional birth-rate distribution using orthogonal distance regression, with effective uncertainties proportional to $\mathcal{R}^{-1/2}$. This weighting emphasizes bins with higher volumetric birth rates, so the fit primarily traces the dominant population.

\subsection{Hypernebula Population}
\label{subsec:HNbpopulation}

\begin{figure*}
\epsscale{1.17}
\plotone{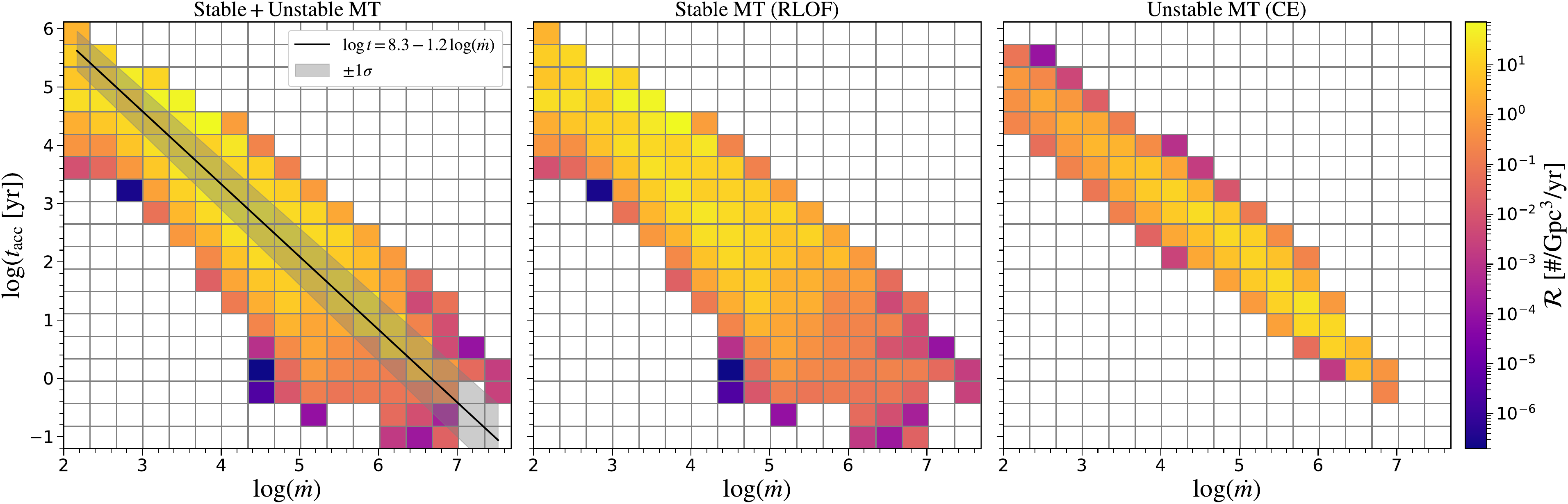}
\caption{
Volumetric rate of hyper-Eddington MT events from our sample in the $\dot{m}$--$t_{\rm acc}$ plane, shown separately for all systems (left), stable RLOF systems (center), and CE systems (right). The black solid line in the left panel shows the best-fit relation for the ridge of the dominant population (Eq.~\eqref{eq:Mdot_tacc_scaling}), illustrating the approximate inverse correlation between the MT rate and the MT duration.
}
\label{fig:rate_2d}
\end{figure*}

\begin{figure*}
\gridline{
\fig{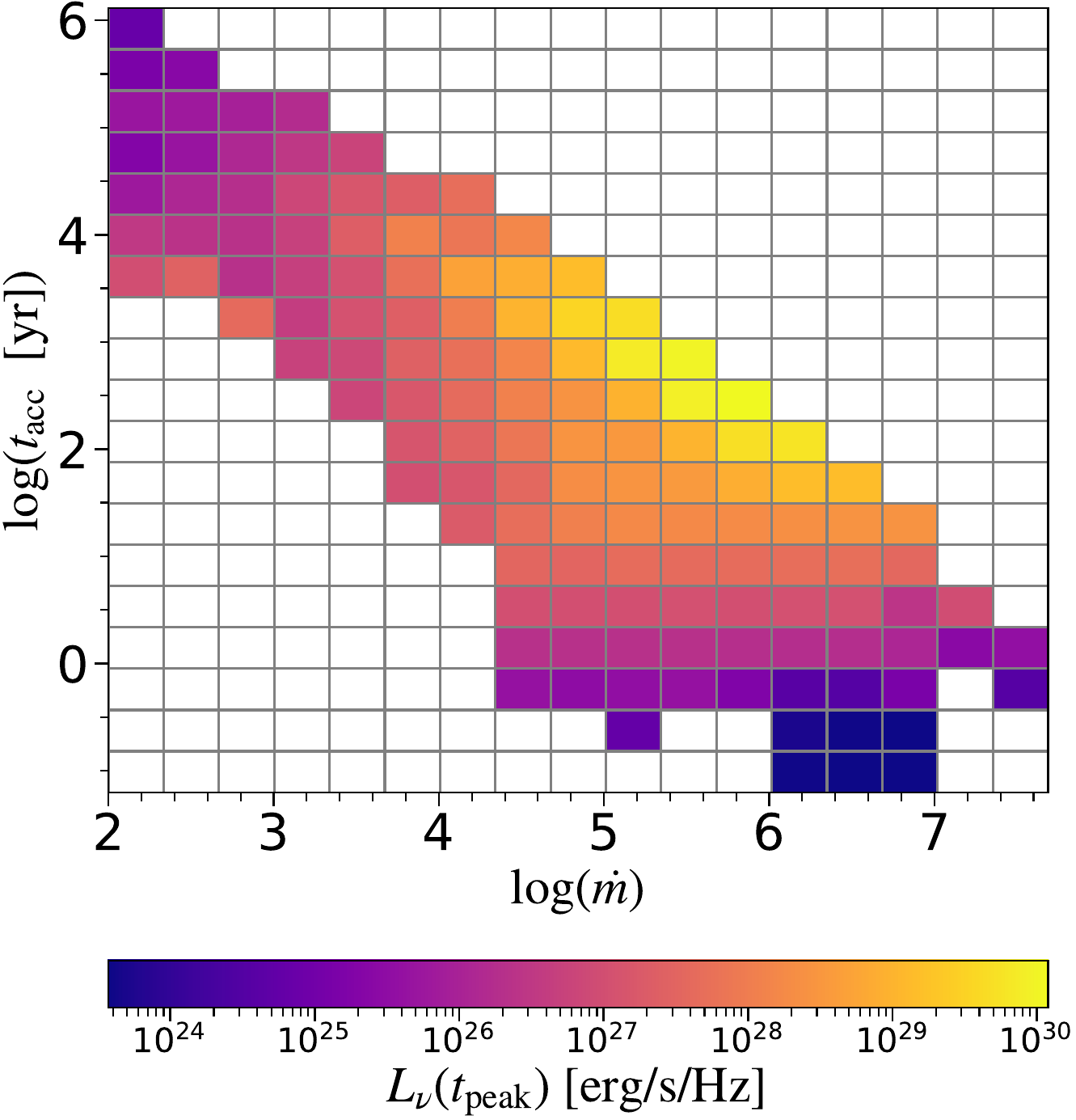}{0.33\textwidth}{(a)  Peak Spectral Luminosity (VLASS)}
\fig{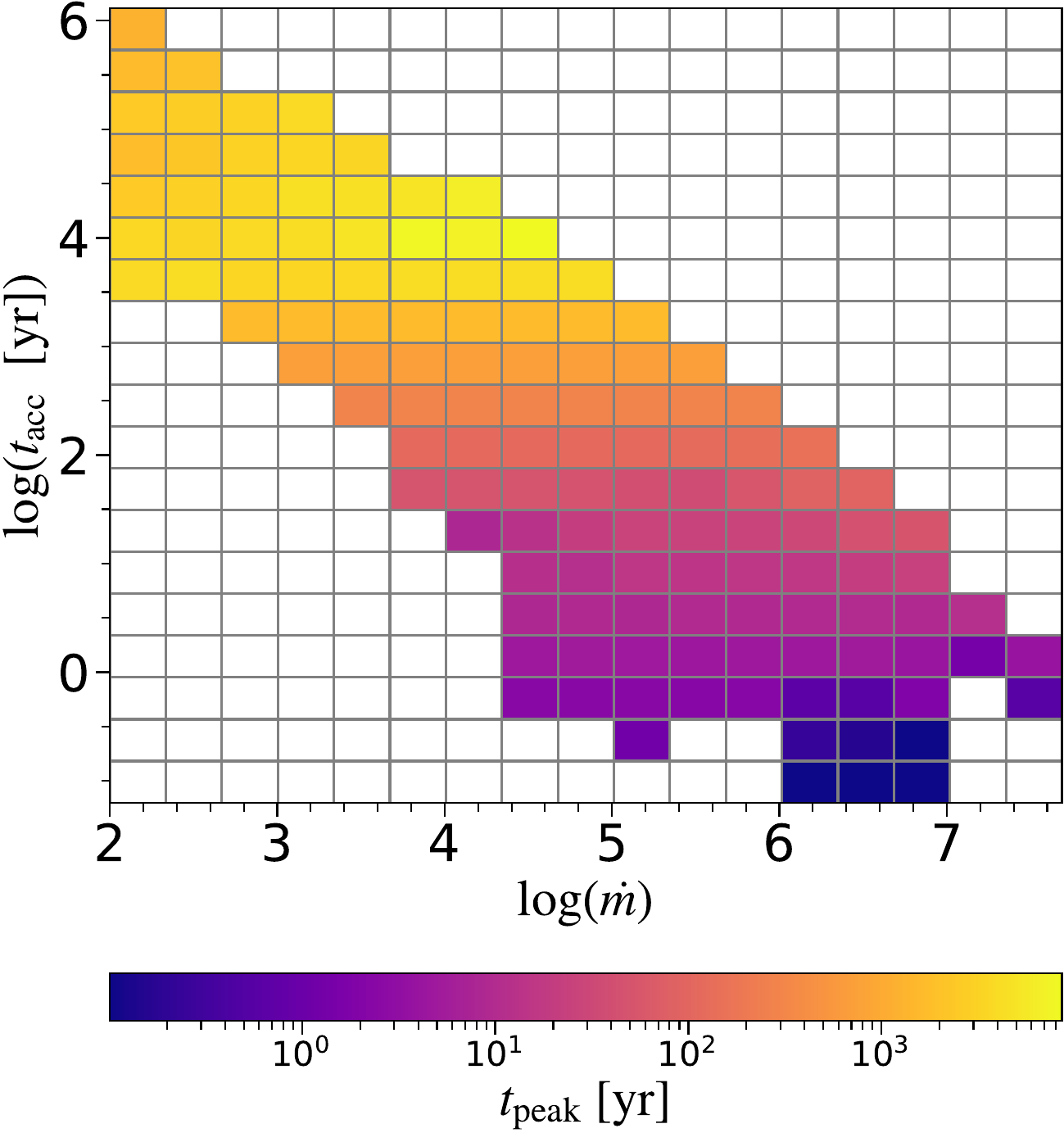}{0.33\textwidth}{(b) Peak Time (VLASS)}
\fig{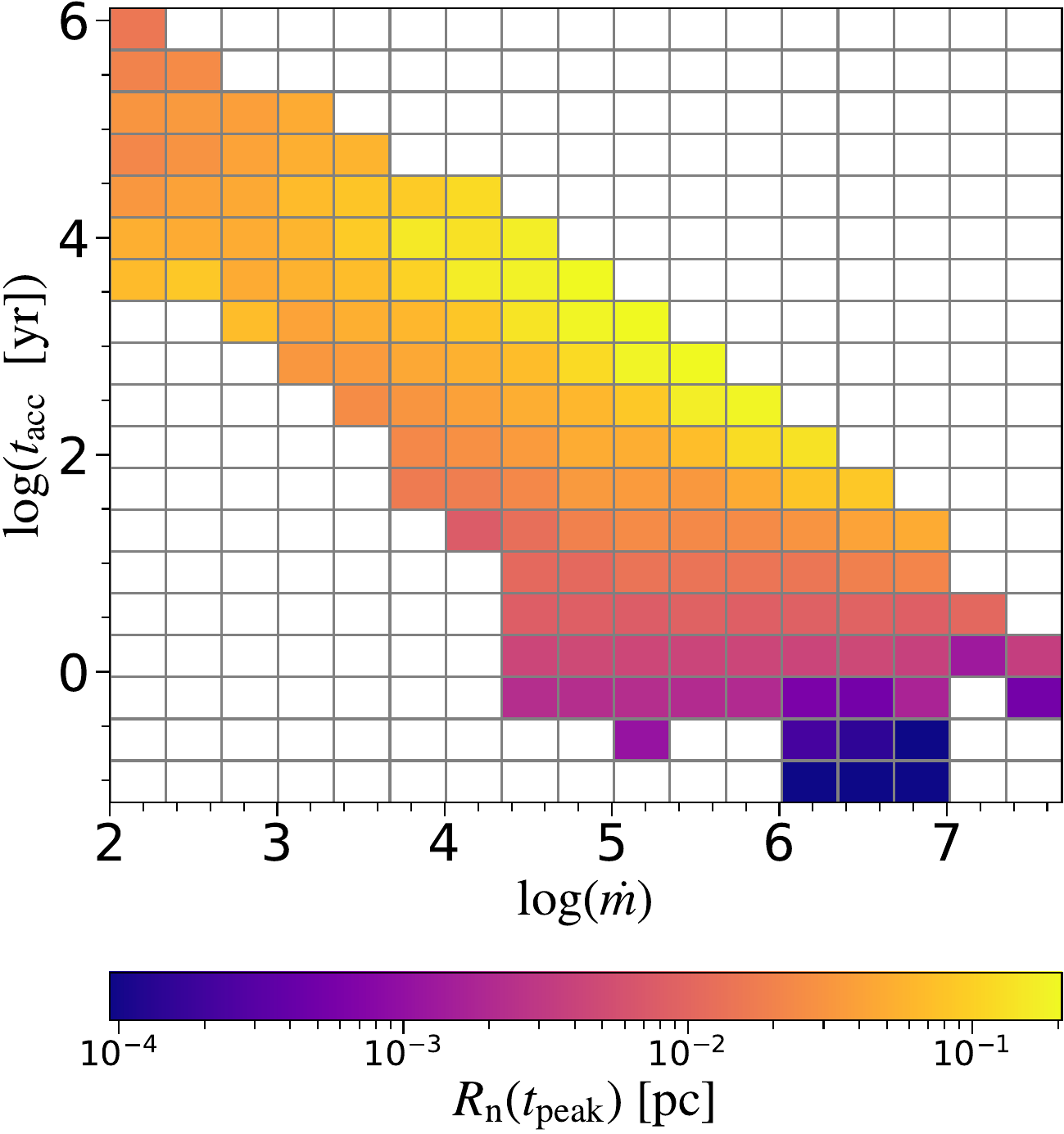}{0.33\textwidth}{(c) Nebula Radius at Peak Time (VLASS)}
}
\caption{
Similar to Fig.~\ref{fig:rate_2d}, but showing several HNb properties for \textsc{Model2}. Separate panels show the peak 3 GHz (VLASS band) radio luminosity, peak timescale, and the nebula radius at the peak time, respectively. Each cell shows the rate-weighted mean value over the MT episodes in that cell.
}
\label{fig:radio_2d}
\end{figure*}

Fig.~\ref{fig:radio_2d} summarizes several key HNb properties across the same parameter space, $\{\dot{m},\, t_{\rm acc}\}$, as in Fig.~\ref{fig:rate_2d}. The different panels show the peak radio luminosity, $L_{\nu}(t_{\rm peak})$, the time to peak, $t_{\rm peak}$, at 3 GHz (the VLASS band), and the nebular radius at the peak, $R_{\rm n}(t_{\rm peak})$. As expected from Figs.~\ref{fig:testcase_lc} and \ref{fig:sed}, longer-lived MT phases (larger $t_{\rm acc}$) generally give rise to longer-lived nebulae. However, the most luminous and largest GHz-emitting nebulae are produced not by the longest-lived systems, but instead by systems with intermediate values of $t_{\rm acc} \sim 10^{2-3}$ yr and $\dot{m} \sim 10^{5-6}$.

\begin{figure*}
\epsscale{1.17}
\plotone{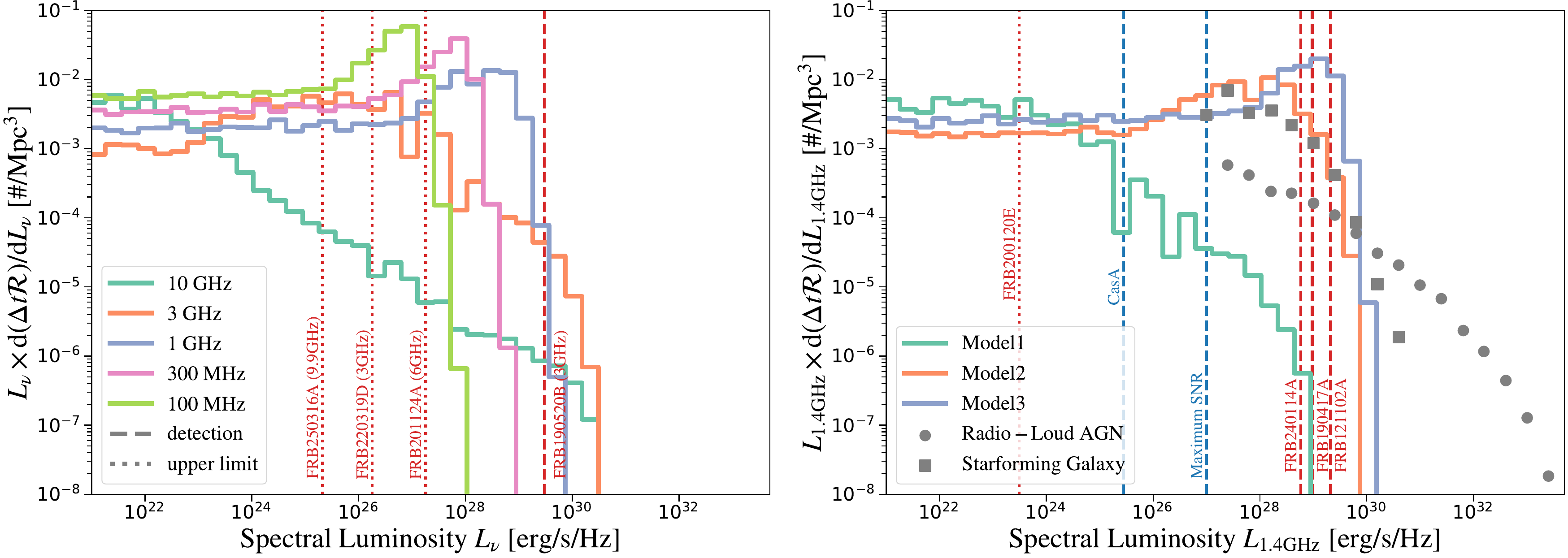}
\caption{
Predicted luminosity function of HNbe in the local universe, obtained by binning each light curve in luminosity and weighting each bin by $\mathcal{R}\,\Delta t$, where $\Delta t$ is the duration spent in that bin (Sec.~\ref{subsec:HNbpopulation}). The left panel shows the result at different observer frequencies for the fiducial \textsc{Model2}, while the right panel shows how the results change for \textsc{Model1}--\textsc{Model3}. For comparison, in the right panel we also show the luminosity functions of radio loud AGN and star-forming galaxies \citep{Mauch&Sadler07}. Vertical lines show the detected (dashed) or upper-limit (dotted) luminosities of radio emission from supernova remnants (e.g., \citealt{Chomiuk&Wilcots09}) as well as the persistent radio sources spatially coincident with FRB sources \citep{Ofek17,Eftekhari+20,Law+22,Niu+22,Bruni_2024_Nature_632_1014,Bruni_2025AA_695L_12, Law_2024_ApJ_967_29L, CHIMEFRB_2025ApJ_989L_48, Kirsten_2022_Nature_602_585, Dong_2024_ApJ_961_44, Bhandari+23, Moroianu+26, Hilmarsson+21}.
}
\label{fig:luminosity_function}
\end{figure*}

Fig.~\ref{fig:luminosity_function} shows the predicted HNb luminosity function in the local universe. We estimate the number of HNbe present at any given time by weighting each source in our sample by its volumetric birth rate and the time interval, $\Delta t$, that it spends in each radio-luminosity bin along its light curve; i.e., the luminosity function is approximated as $L_{\nu}(d(\Delta t \mathcal{R})/dL_{\nu})$. 
For comparison, we also show the luminosity functions of radio-loud AGN and star-forming galaxies \citep{Mauch&Sadler07}, as well as the luminosities of supernova remnants \citep{Chomiuk&Wilcots09} and of the persistent radio sources associated with FRBs \citep{Ofek17,Eftekhari+20,Law+22,Niu+22,Dong+24}. We find that the HNbe population spans a broad range of radio luminosities, extending up to $L_{\nu} \sim 10^{28-30}$ erg s$^{-1}$ Hz$^{-1}$ at 1.4 GHz, depending sensitively on the adopted HNb model. We return to the implications of this population in Sec.~\ref{sec:discussion}.

\subsection{Radio Survey Detection Prospects} 
\label{subsec:surveydetect}

A luminosity function akin to Fig.~\ref{fig:luminosity_function} is a useful observable in an idealized volume-limited survey with proper cosmological corrections. In practice, however, most radio surveys are flux-limited, and redshift information for the source is often incomplete for much of the source population. Therefore we adopt a forward-modeling approach to estimate the detectable HNb population in a blind radio survey.

\subsubsection{Method}
\label{subsubsec:rates}

The number of HNbe detectable in a given radio survey, $\mathcal{N}_{\mathrm{det}}$, depends on the volumetric birth rate of the source population, $\mathcal{R}(z)$ (Sec.~\ref{sec:rates}), the rest-frame radio light curve, $L_{\nu,\mathrm{rest}}(\nu_{\mathrm{rest}}, t_{\mathrm{rest}})$, and the survey limiting sensitivity, $F_{\nu,\mathrm{lim}}$. For a source at redshift $z$, the observed flux density at observing frequency $\nu_{\mathrm{obs}}$ and observer-frame time $t_{\mathrm{obs}}$ is
\begin{equation}
    F_{\nu,\mathrm{obs}}(\nu_{\mathrm{obs}}, t_{\mathrm{obs}}; z)
    =
    \frac{(1+z)\,
    L_{\nu,\mathrm{rest}}\!\left[(1+z)\nu_{\mathrm{obs}},\, t_{\mathrm{obs}}/(1+z)\right]}
    {4\pi D_{L}(z)^2},
    \label{eq:fnu_obs}
\end{equation}
where $D_L(z)$ is the luminosity distance. The rest-frame and observer-frame quantities are expressed as follows:
\begin{equation}
    \nu_{\mathrm{rest}} = (1+z)\nu_{\mathrm{obs}},
    \quad
    t_{\mathrm{obs}} = (1+z)t_{\mathrm{rest}}.
\end{equation}
When $F_{\nu,\mathrm{obs}} > F_{\nu,\mathrm{lim}}$, the source is detectable. We define the detection horizon, $z_{\mathrm{max}}$, as the maximum redshift at which the source is detectable at any observer-frame time, and the observer-frame detectable duration, $\Delta t_{\mathrm{obs}}(z)$.

The total number of detectable sources in a survey covering solid angle $\Omega$ is
\begin{equation}
    \mathcal{N}_{\mathrm{det}}(z)
    =
    \frac{\Omega}{4\pi}
    \int_0^{z<z_{\max}}
    \mathcal{R}(z)\,
    \frac{\Delta t_{\mathrm{obs}}(z)}{1+z}\,
    \frac{dV_c}{dz}\,
    dz,
    \label{eq:Ndet_cosmo}
\end{equation}
where $dV_c/dz$ is the differential comoving volume over the full sky. We adopt the cosmological parameters of $H_0 = 73.2$ km s$^{-1}$ Mpc$^{-1}$ \citep{Burns_2018_ApJ_869_56B} and $\Omega_{\rm m} = 0.3$, and integrate Eq.~\eqref{eq:Ndet_cosmo} up to $z = 1.5$. The factor $(1+z)^{-1}$ converts the observer-frame detectable duration to the rest-frame duration.

\begin{deluxetable*}{ccccccc}
\tablecaption{Radio Survey Properties \label{tab:survey}}
\tablewidth{700pt}
\tabletypesize{\scriptsize}
\tablehead{
\colhead{Survey} & \colhead{Central Frequency} & \colhead{Sensitivity (5$\sigma$)} & \colhead{Angular Resolution} & \colhead{Sky Coverage} & \colhead{Cadence} & \colhead{Reference} \\
& [{\rm GHz}] &  [mJy] & [arcsec] & [${\mathrm{deg^2}}$] & & 
} 
\startdata
VLA/VLASS & $3.0$ & 0.7 & $\approx 2$ & $\approx 34,000$ & $\approx 2.5 {~\mathrm{years}}$ & 1 \\ 
LOFAR/LoTSS & $0.14$ & 0.42 & $\approx 6$ & 18,000 & -- & 2, 3, 4 \\
ASKAP/VAST & $0.89$ & 1.2 & $\approx 12-20$ & $\approx 12,000$ & days - months & 5, 6 \\
DSA-2000 & $1.0$ & 0.01 & $\approx 3.5$ & $\approx 31,000$ & $\approx 4 {~\mathrm{months}}$ & 7 \\
\enddata
\tablerefs{(1) \citet{Lacy+20}; (2) \citet{Shimwell_2017_AA_598A_104S}; (3) \citet{Shimwell2022AA}; (4) \citet{Shimwell_2026_AA_707A_198S}; (5) \citet{Murphy2021PASA}; (6) \citet{deRuiter_2026_PASA_43_39D}; (7) \citet{Hallinan2019BAAS}.}
\end{deluxetable*}

\subsubsection{Survey Detected Population}

For the existing or planned radio surveys summarized in Table~\ref{tab:survey}, we calculate the number of detectable HNbe, $\mathcal{N}_{\mathrm{det}}$, along with other properties of the detected population. 

\begin{figure*}
\epsscale{1.17}
\plotone{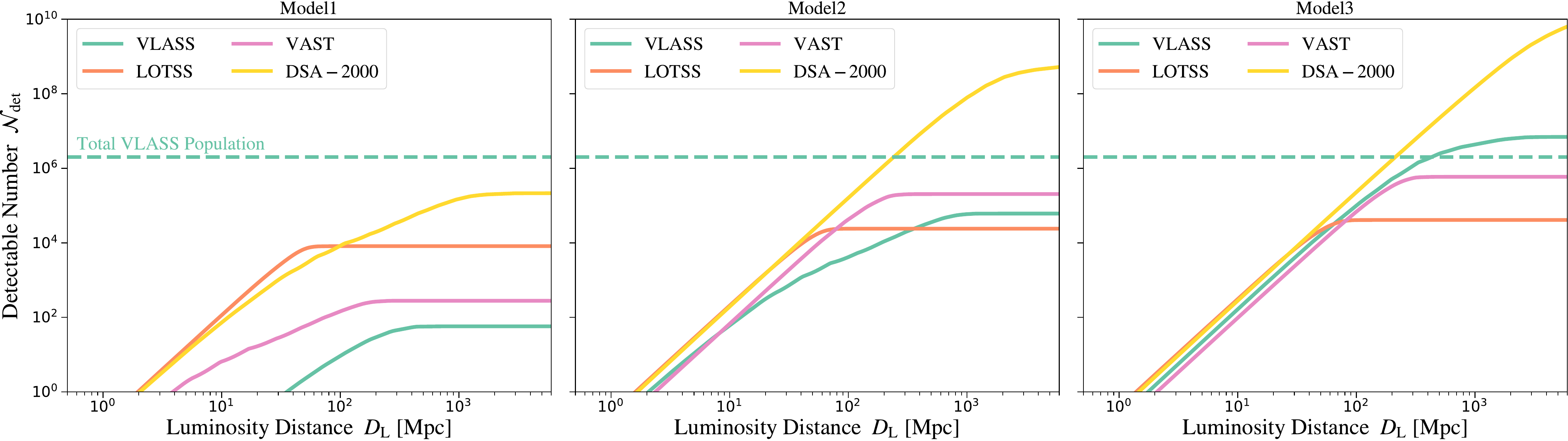}
\caption{
Cumulative number of HNb sources up to a given luminosity distance for various radio surveys as marked (see Tab.~\ref{tab:survey}). Panels denote results for \textsc{Model1}, \textsc{Model2}, and \textsc{Model3}, respectively.
}
\label{fig:cumulative_models}
\end{figure*}

Fig.~\ref{fig:cumulative_models} shows the cumulative number of detectable HNbe out to a given luminosity distance, for each of the three light-curve models (Tab.~\ref{tab:modelparams}). The detectable sample increases with the observed volume, until reaching its maximum near the detection horizon of the most luminous sources in the sample. Unsurprisingly, the large \'{e}tendue of DSA-2000 results in the largest total number of detections for all of the models. For \textsc{Model1}, LOFAR/LoTSS also detects $\sim 10^4$ HNbe, two orders of magnitude more than VLASS, because its low-frequency band is better suited to the SED of \textsc{Model1} (which peaks at lower frequency because of the lower jet velocity). On the other hand, for the most optimistic model (\textsc{Model3}), the total VLASS source count of $\sim 7\times 10^{6}$ HNbe exceeds the total number of radio sources detected thus far in the entire survey of $\sim 2\times 10^{6}$ \citep{Lacy+20}. This suggests that our highest $v_{\rm j} = 0.25c$ model likely greatly overestimates the luminosity of the MT population.

\begin{figure*}
\epsscale{1.17}
\plotone{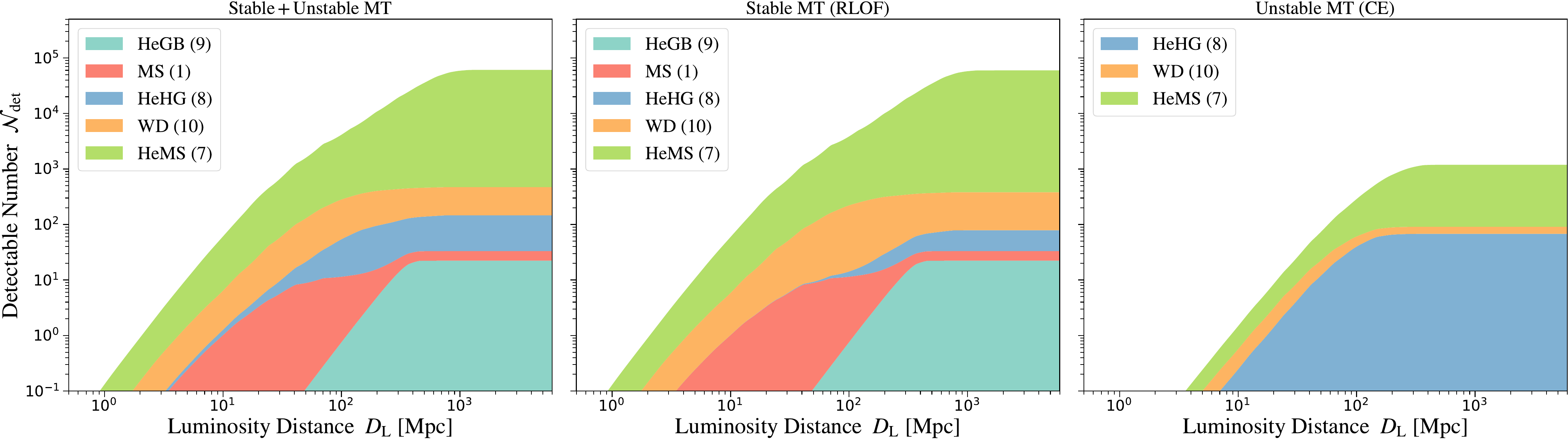}
\caption{
Cumulative number of detectable HNbe up to a given luminosity distance by VLASS, broken down by the final state of the MT donor star for our \textsc{Model2}. Left panel shows all channels, while center and right panels show separately stable RLOF and unstable CE channels, respectively.
}
\label{fig:cumulative_models2}
\end{figure*}

Fig.~\ref{fig:cumulative_models2} breaks down the detected VLASS source population for \textsc{Model2} into the contributions from different donor states.  We see that MT systems leaving stars on the helium main-sequence (HeMS) as the final donor state dominate the detected population. This is expected because a plurality of our hyper-Eddington MT events are of the type HG$\rightarrow$HeMS (Fig.~\ref{fig:beforeaftermapping}), and these are precisely the systems characterized by intermediate values of $t_{\rm acc}$ and $\dot{m}$ that give rise to the longest-lasting and most luminous HNbe at GHz frequencies (Fig.~\ref{fig:radio_2d}).

\section{Discussion: Prospects for Identification} \label{sec:discussion}

The central challenge to uniquely identifying an HNb population is distinguishing these systems from many other populations of unresolved synchrotron sources found in modern wide‐field radio surveys. This section explores several observable properties that could help separate HNbe from these other sources, thus enabling their characterization. We also explore the potential connections between HNbe and other astrophysical transients.

\subsection{Comparison with Other Radio Sources}

A potentially distinguishing property of HNbe is their luminosity function (Fig. \ref{fig:luminosity_function}). The predicted HNb luminosity functions are flatter than those of radio-loud AGN and star-forming galaxies \citep{Mauch&Sadler07} and are truncated at a model-dependent maximum luminosity. Near this cutoff, \textsc{Model3} exceeds the observed local radio luminosity function, suggesting that our highest-$v_{\rm j}$ model overestimates the HNb luminosities (see also Sec.~\ref{subsec:surveydetect}). HNbe can also be substantially more luminous than supernova remnants \citep[$L_{\nu} \lesssim 10^{27}\,\mathrm{erg\,s^{-1}\,Hz^{-1}}$;][]{Chomiuk&Wilcots09}, and their luminosities overlap with those of the persistent radio sources associated with some repeating FRBs \citep[see also Fig.~\ref{fig:luminosity_function};][]{Chatterjee+17,Ofek17,Eftekhari+20,Law+22,Niu+22,Bhandari+23,Dong+24,Ibik+24,Snelders+25,Moroianu+26}, which may motivate a physical association between repeating FRBs and MT binaries (\citetalias{Sridhar2022ApJ}).

\begin{figure}
\centering
\epsscale{1.17}
\plotone{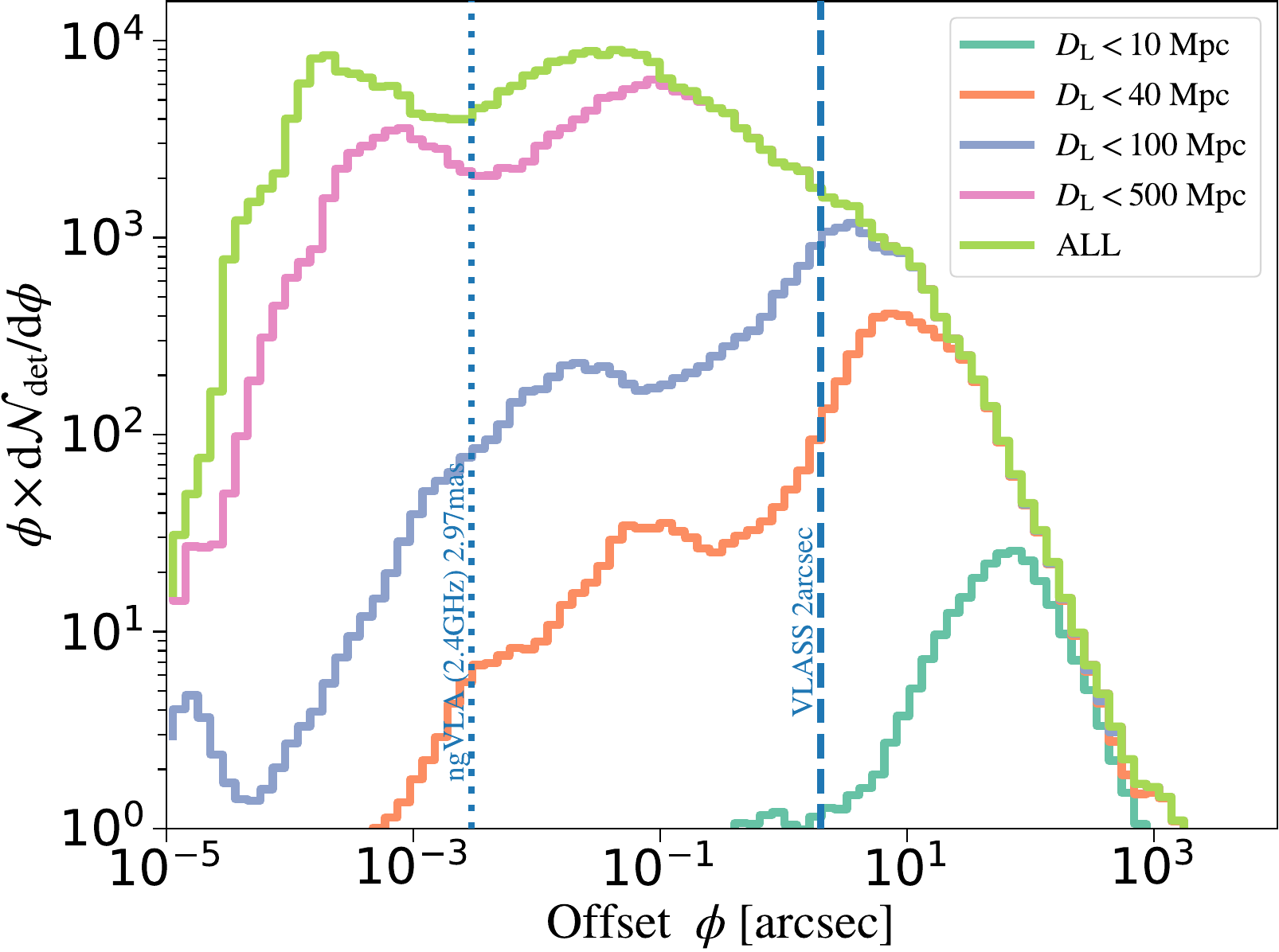}
\caption{
Distribution of the predicted angular offsets ($\phi$) of HNbe from their birth sites as a result of the supernova kick responsible for birthing the accreting compact object, as calculated for \textsc{Model2} at $3$ GHz.  Colored lines show the distributions for different assumed limiting distances of the source population as marked in the legend. Vertical blue lines show the estimated angular resolutions of VLASS and ngVLA.
}
\label{fig:offset}
\end{figure}

Another observable property of HNbe is their spatial locations within their host galaxies. AGN generally coincide with the centers of galaxies, though there are exceptions for the lowest-mass active galaxies or in the case of wandering supermassive black hole candidates (e.g., \citealt{Reines+20}).  Massive stellar binaries are born in star-forming regions, but some offset from their birth-sites is expected during the lifetime of HNbe.  Fig.~\ref{fig:offset} shows the distribution of angular offsets of HNbe from their birth sites due to the kick the binary receives from the first supernova (which births the accreting compact object), calculated for our detected 3 GHz VLASS sample using \textsc{Model2}. We estimate the maximum angular offset as $\phi \approx v_{\rm sys}\, t_{\rm SN1\to HNb}/D_{\rm A}$, where $v_{\rm sys}$ is the post-SN systemic velocity of the binary from \textsc{COMPAS} (typically $\sim 10$–$20$ km s$^{-1}$), $t_{\rm SN1\to HNb}$ is the time from the first SN to the HNb phase, and $D_{\rm A}$ is the angular diameter distance. We see that angular offsets of $\sim 1-10$ arcsec at $D_{\mathrm{A}} \approx 100$ Mpc are typical, which in general should be sufficient to separate HNbe from their star-forming regions.

\begin{figure}
\centering
\epsscale{1.17}
\plotone{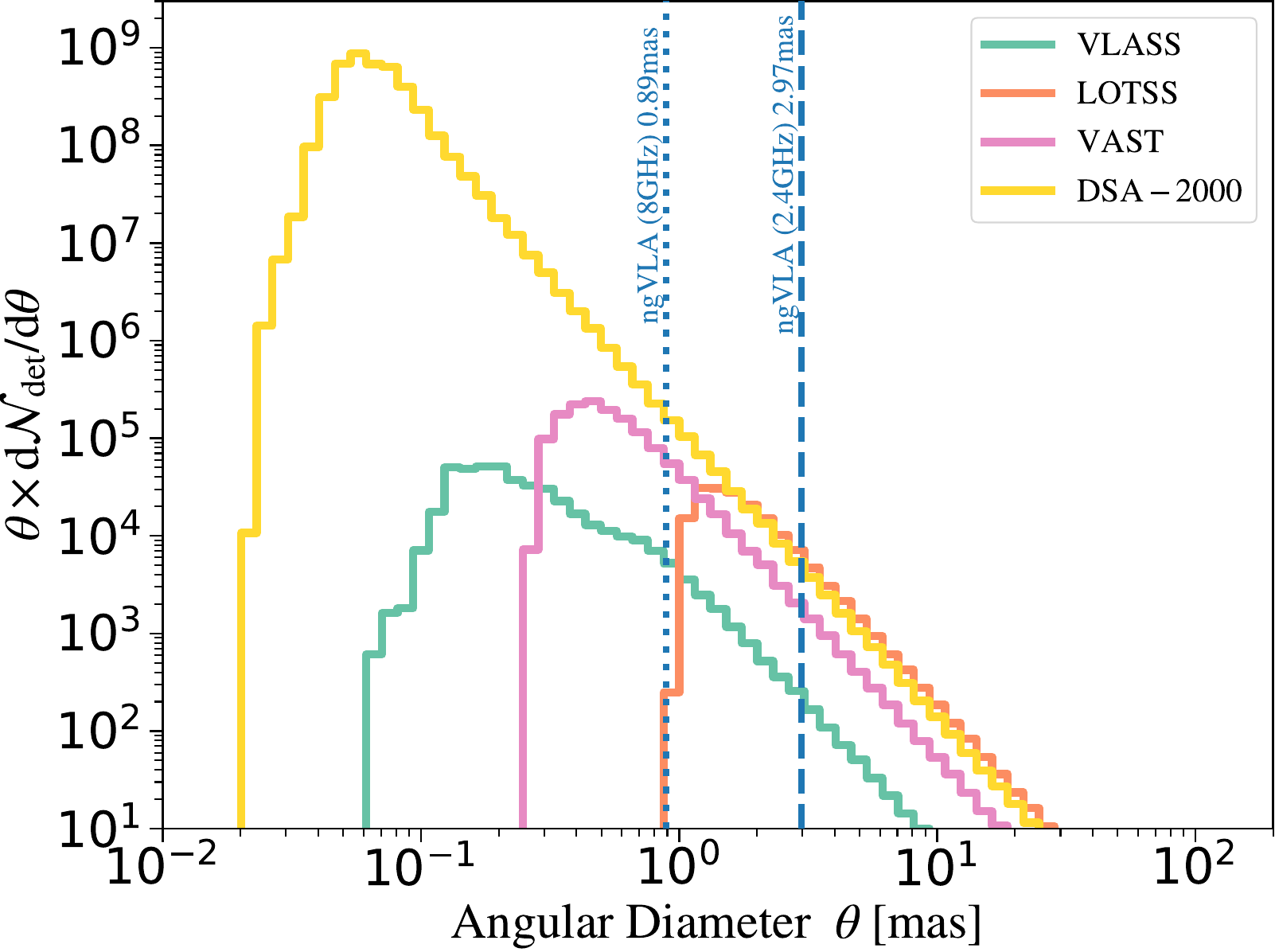}
\caption{
Distribution of angular diameter size of HNbe ($\theta$) at the peak time of their light curves, as detected by various radio surveys based on \textsc{Model2}. Shown for comparison with vertical blue lines are the estimated resolution of the next-generation VLA (ngVLA) at 2.4 GHz and 8 GHz, respectively \citep{Murphy+18}. Although the vast majority of the detected sources are too far away, a sizable number of nebulae are potentially close enough to resolve.
}
\label{fig:dNdet_theta}
\end{figure}

A fraction of nearby HNbe will be sufficiently large to have their radio emission region spatially resolved. The diameter of the nebula at peak luminosity is typically less than a parsec (Fig.~\ref{fig:radio_2d}), comparable in size to a young supernova remnant but generally more compact than radio AGN. Fig.~\ref{fig:dNdet_theta} shows the predicted angular diameters of our survey-detected HNbe population, which we calculate as $\theta \approx 2R_{\rm n}(t_{\rm peak})/D_{\rm A}$.  Depending on the survey and frequency of observation, tens of HNbe should have sufficient sizes $\theta \gtrsim 10$ mas to be resolved by VLBI.  Although VLBI requires a higher flux density than detection alone, nebulae with the largest angular sizes are also generally the closest and brightest sources.

\begin{figure}
\centering
\epsscale{1.17}
\plotone{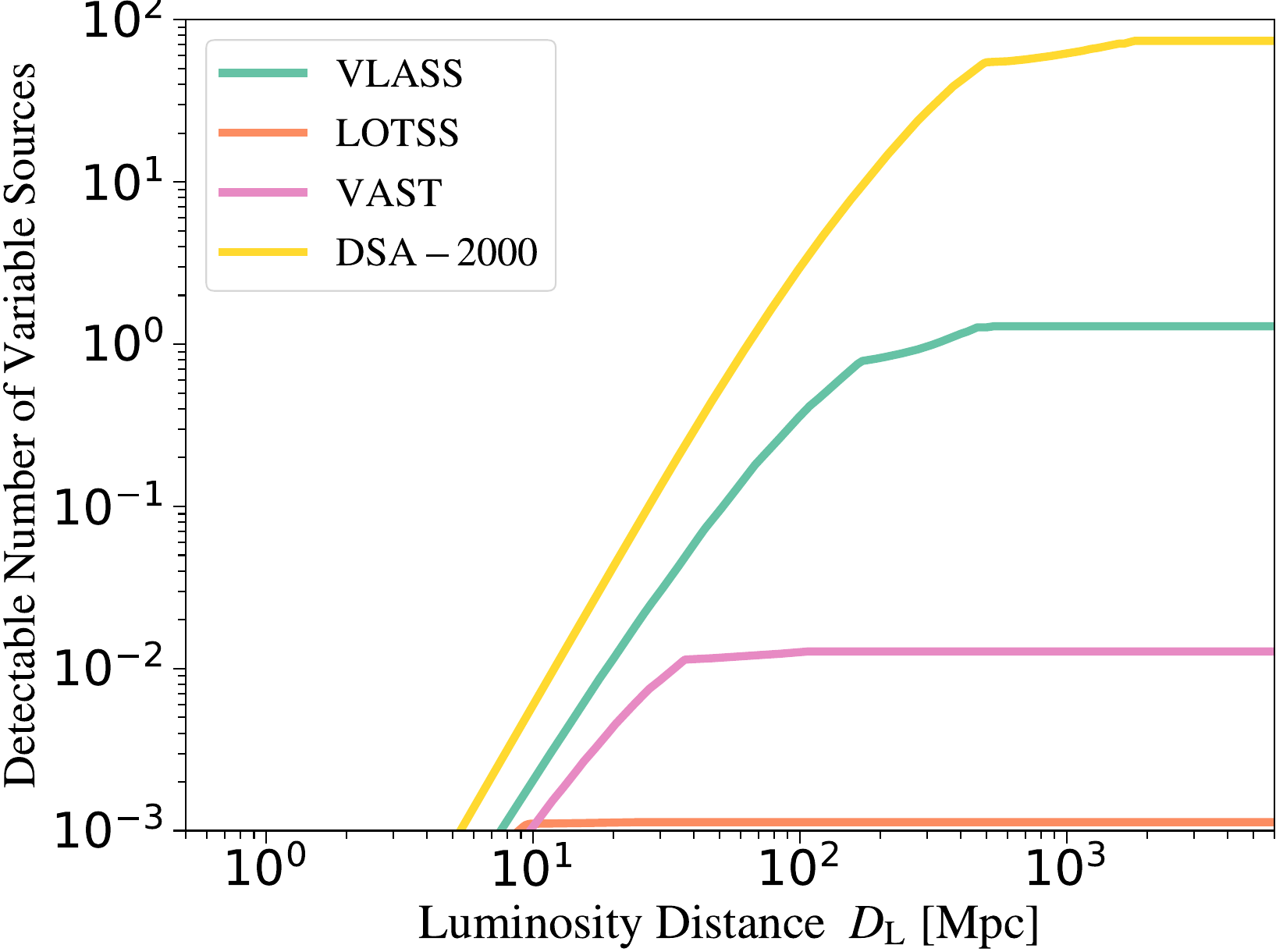}
\caption{
Number of time-varying HNbe detected by various radio surveys as a function of the source distance in \textsc{Model2}. We define time-varying nebulae as those for which the flux varies by a factor of 3 over a 20-year baseline.
}
\label{fig:variable_source}
\end{figure}

Young orphan GRB afterglows can reach radio luminosities comparable to those of HNbe, with estimated volumetric rates of $\sim 10^{3}$ Gpc$^{-3}$ yr$^{-1}$ \citep{Law+18}. However, such afterglows typically evolve on timescales of years to decades, whereas most HNbe are expected to remain stable over much longer durations, with characteristic accretion timescales of $t_{\rm acc} \sim 10^{3-5}$ yr. This large difference in variability timescales offers a useful way to distinguish between the two populations. To assess the fraction of HNbe that could show time variability over the duration of a typical radio survey, we analyzed the light curve model using a method similar to our detectable number estimate. We weighted each system by the fraction of its detectable lifetime over which the luminosity varies by more than a factor of three over a 20-year baseline, following the same criterion used in \citetalias{Sridhar2022ApJ}. Fig.~\ref{fig:variable_source} shows the number of such variable HNbe detectable by each survey. Although the fraction is small, we find that a few variable sources may be detected with VLASS, and up to $\sim 100$ with DSA--2000. Long-baseline monitoring thus offers a key tool to distinguish HNbe from other explosive transients. Therefore, HNbe occupy a unique parameter space -- physically small but luminous persistent sources with spatial offsets -- in the radio detected population.

Another potentially distinguishing feature of HNbe is the shape of their radio SED, which in our model arises from a combination of the assumed thermal distribution of the emitting electrons and synchrotron-self-absorption (SSA) effects. In particular, when the observing frequency lies below the SSA peak, the spectrum is optically thick and follows $F_{\nu} \propto \nu^{2}$, although in detail it depends on the source geometry and electron distribution.  On the other hand, observing frequencies well above the SSA peak will observe an optically-thin spectrum ($\alpha \gtrsim 2$, where $F_{\nu} \propto \nu^{-\alpha}$). This contrasts with other compact radio sources: GRB afterglows typically show optically-thin synchrotron emission with $\alpha \lesssim 1.5$ \citep[e.g.,][]{Sari_1998_ApJ_497_L17S}, while radio AGNs often exhibit flatter or even inverted spectra \citep[e.g., $\alpha \lesssim 0.7$;][]{Condon_1992_ARAA_30_575C}. Thus, radio sources observed at different frequencies (e.g., by distinct surveys or via follow-up observations of candidate sources) could help identify a promising HNbe candidate list. 

However, we repeat our earlier caution that our predictions for the HNbe spectral shape are sensitive to several uncertainties in our models, particularly the assumption of relativistic Maxwellian electrons into the nebula.  Other physical effects, such as time-dependent or spatially-varying shock structures, or additional emission from non-thermal electrons accelerated by shocks or magnetic reconnection, are likely to substantially broaden the spectrum and may render it more degenerate with other source populations.

\subsection{Connection to Other Transients}

For a small fraction of HNbe discovered by radio surveys, the donor star may be sufficiently luminous to detect directly.  Fig.~\ref{fig:hr} shows that the most luminous donor stars are those on the upper HeHG branch located in the upper right hand corner of the HR diagram. For these systems (mainly FGB/EAGB$\rightarrow$HeHG), the MT timescale is typically short ($\lesssim 10^{3-4}$ yrs). As a result, their HNbe phases are challenging to detect and represent only a modest fraction of the total observed population (blue region in Fig.~\ref{fig:cumulative_models2}). Nevertheless, assuming the $\lesssim 100$-Mpc horizon within which sensitive space telescopes (e.g., {\it Roman} and {\it JWST}) could detect and potentially spatially resolve such sources, we predict a few to tens of radio-detected FGB/EAGB$\rightarrow$HeHG systems. 

Recently, for one of the closest and best localized FRB 20250316A ($\sim 40 {\rm ~Mpc}$; \citealt{CHIMEFRB_2025ApJ_989L_48}), follow-up observations by the {\it JWST} \citep{Blanchard_2025_ApJ_989_L49} detected a luminous infrared point source consistent with the FRB location.  Assuming this source is physically associated with the FRB, its luminosity is consistent with being a HeHG/HeGB star.  Although this could support an association between FRB and MT binary systems \citep{Sridhar+21a}, no persistent radio emission was detected either in archival pre-burst VLA imaging ($L_\nu < 3\times10^{25}$ erg s$^{-1}$ Hz$^{-1}$ at 8–10 GHz; \citealt{Golay+25}) or in follow-up observations ($L_\nu < 2.1\times10^{25}$ erg s$^{-1}$ Hz$^{-1}$ at 9.9 GHz; \citealt{CHIMEFRB_2025ApJ_989L_48}).  These upper limits constrain any associated HNb emission, although assessing their compatibility with the proposed MT-binary interpretation requires accounting for the nebular age and the sensitivity of the high-frequency radio luminosity to the assumed jet properties.

Finally, it is of interest to consider whether HNbe emission could be detected as precursor emission to energetic transients associated with binary evolution. For example, the progenitors of many stripped-envelope SNe may lose their envelopes through MT events \citep[e.g.,][]{Wheeler+75,Langer12,Eldridge+13,Smartt15,Lu+23}. In such systems, a key question is whether the HNb phase remains active at the second SN explosion\footnote{A related possibility is that the ejecta of the second SN interact with the HNb shell, producing an interacting SN. We do not attempt to model this scenario in this paper, because the identification of such events from the \textsc{COMPAS} outputs is highly uncertain. The observed properties of interacting SNe depend sensitively on the CSM mass, the SN ejecta properties, and the efficiency of converting kinetic energy into radiation, which are not robustly constrained in our models. In addition, their photometric properties span a wide range of luminosities and timescales, making their classification based only on photometry difficult.}.

\begin{figure*}
\gridline{
    \fig{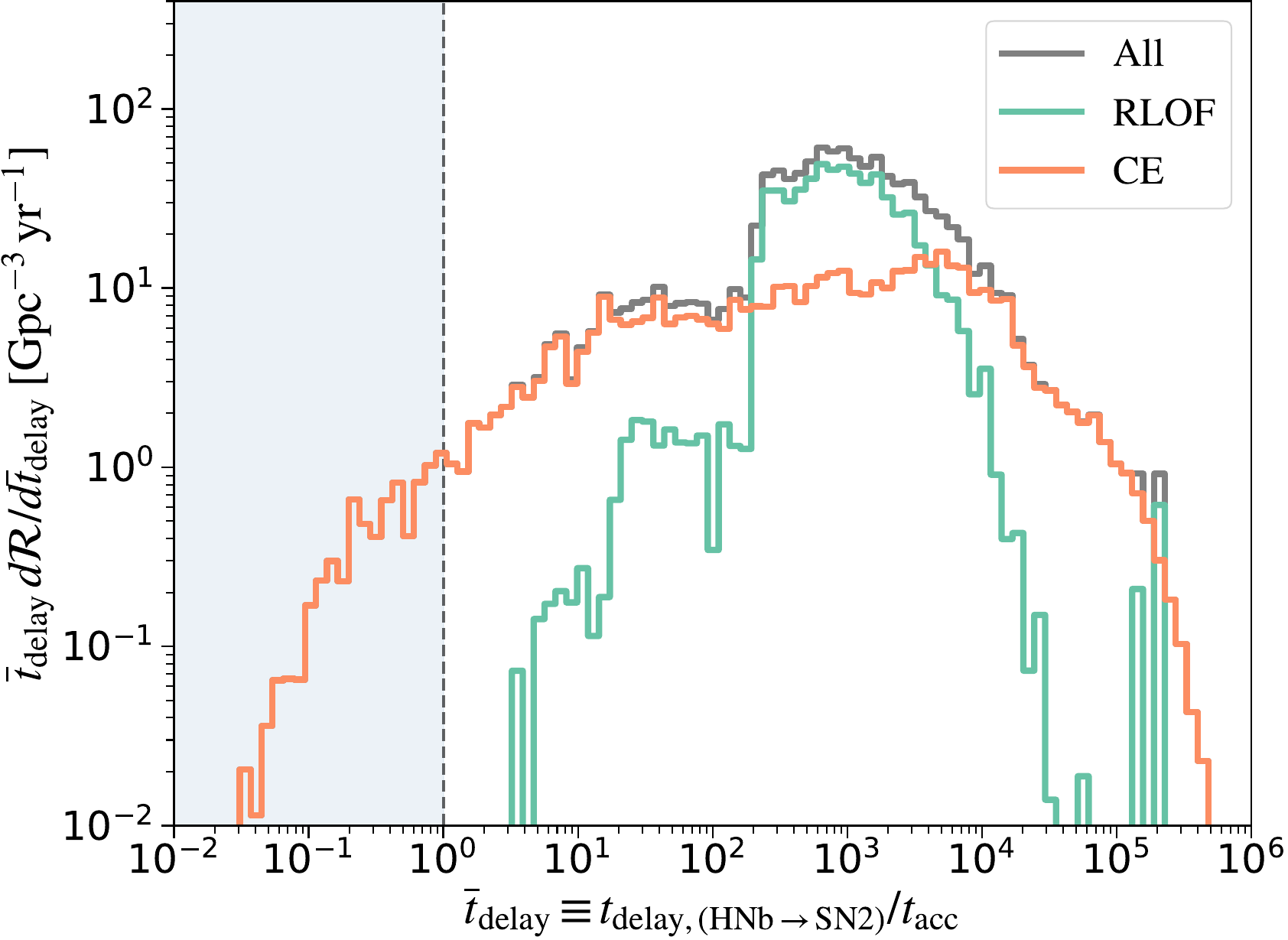}{0.49\textwidth}{(a) Delay time distribution for HNBe}
    \fig{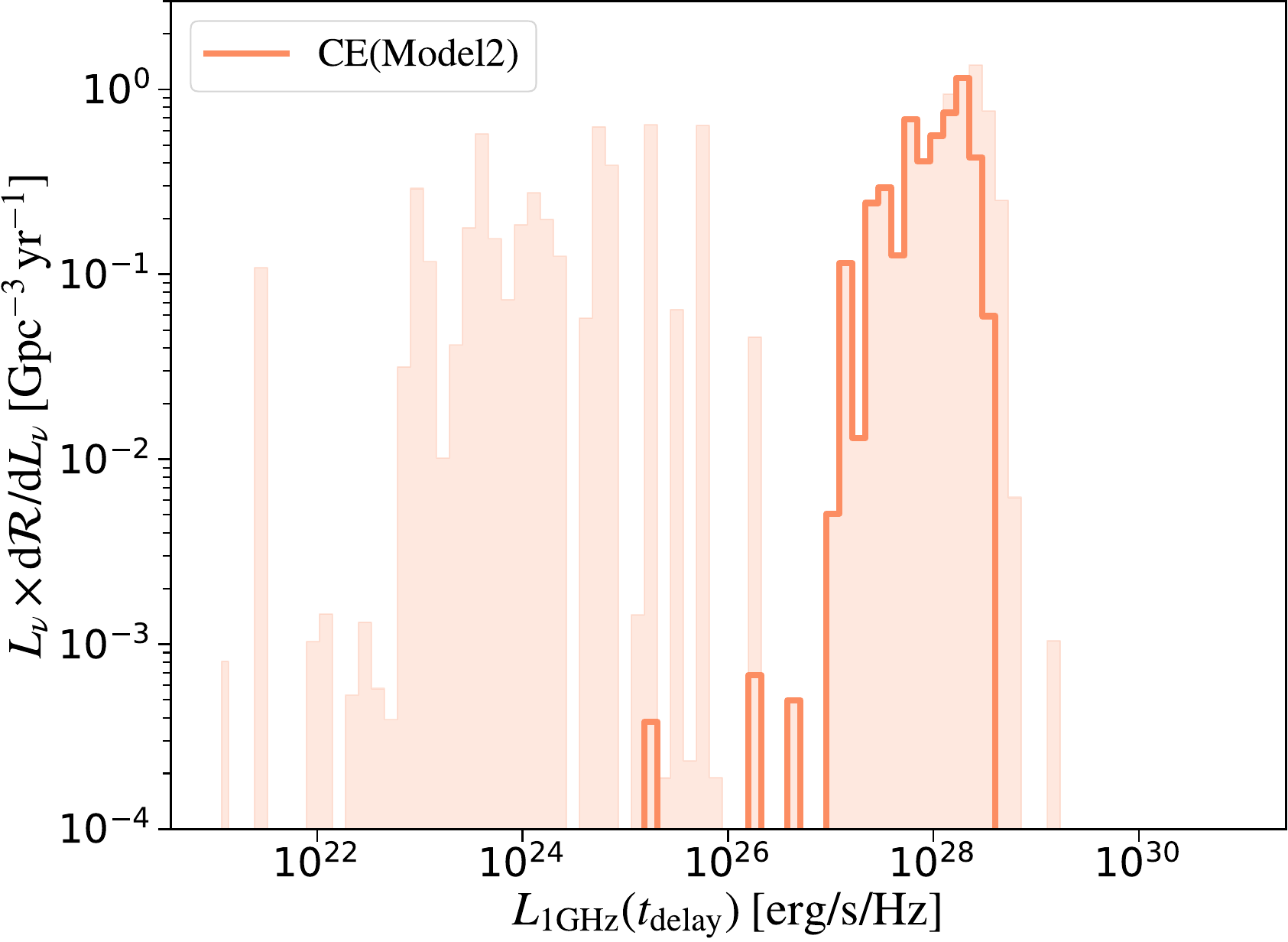}{0.49\textwidth}{(b) Luminosity functions for CE systems}
}
\caption{
(a): Rate-weighted delay ratio distribution, $\bar{t}_\mathrm{delay} \equiv t_\mathrm{delay}/t_\mathrm{acc}$ for all HNb systems undergoing a second SN (SN2) at $z\approx 0$ in \textsc{Model2}, where $t_\mathrm{delay}$ is the time from the onset of the HNb phase to the second SN. The blue shaded region ($\bar{t}_\mathrm{delay} < 1$) marks systems in which HNbe is still active at the time of the second SN. Curves show the differential distribution $\bar{t}\,d\mathcal{R}/d\bar{t}$ for all systems (gray), stable RLOF (green), and CE (orange) channels.
(b): Rate-weighted radio luminosity function $L_\nu\,d\mathcal{R}/dL_\nu$ at 1~GHz for CE systems with $\bar{t}_\mathrm{delay} < 1$, evaluated at $t = t_\mathrm{delay}$. The shaded band spans three parameter models, and the solid line shows the estimate based on \textsc{Model2}.
}
\label{fig:delaytime}
\end{figure*}

To address this, Fig.~\ref{fig:delaytime}(a) shows the distribution of the delay time distribution, $\bar{t}_\mathrm{delay} \equiv t_\mathrm{delay}/t_\mathrm{acc}$, where $t_\mathrm{delay}$ is the interval between the onset of the HNb phase and the second SN. Systems with $\bar{t}_\mathrm{delay} < 1$ correspond to cases in which the HNb phase is still active at the time of the second explosion. As shown in Fig.~\ref{fig:delaytime}(a), only a small fraction, $\sim 1\%$, of HNbe systems undergoing a second SN satisfy this condition, and these are dominated entirely by the CE channel. Fig.~\ref{fig:delaytime}(b) further shows the radio luminosity function of these active systems, evaluated at $t=t_\mathrm{delay}$, demonstrating that only the bright population would remain observable by the time of explosion.

\begin{figure*}
\gridline{
    \fig{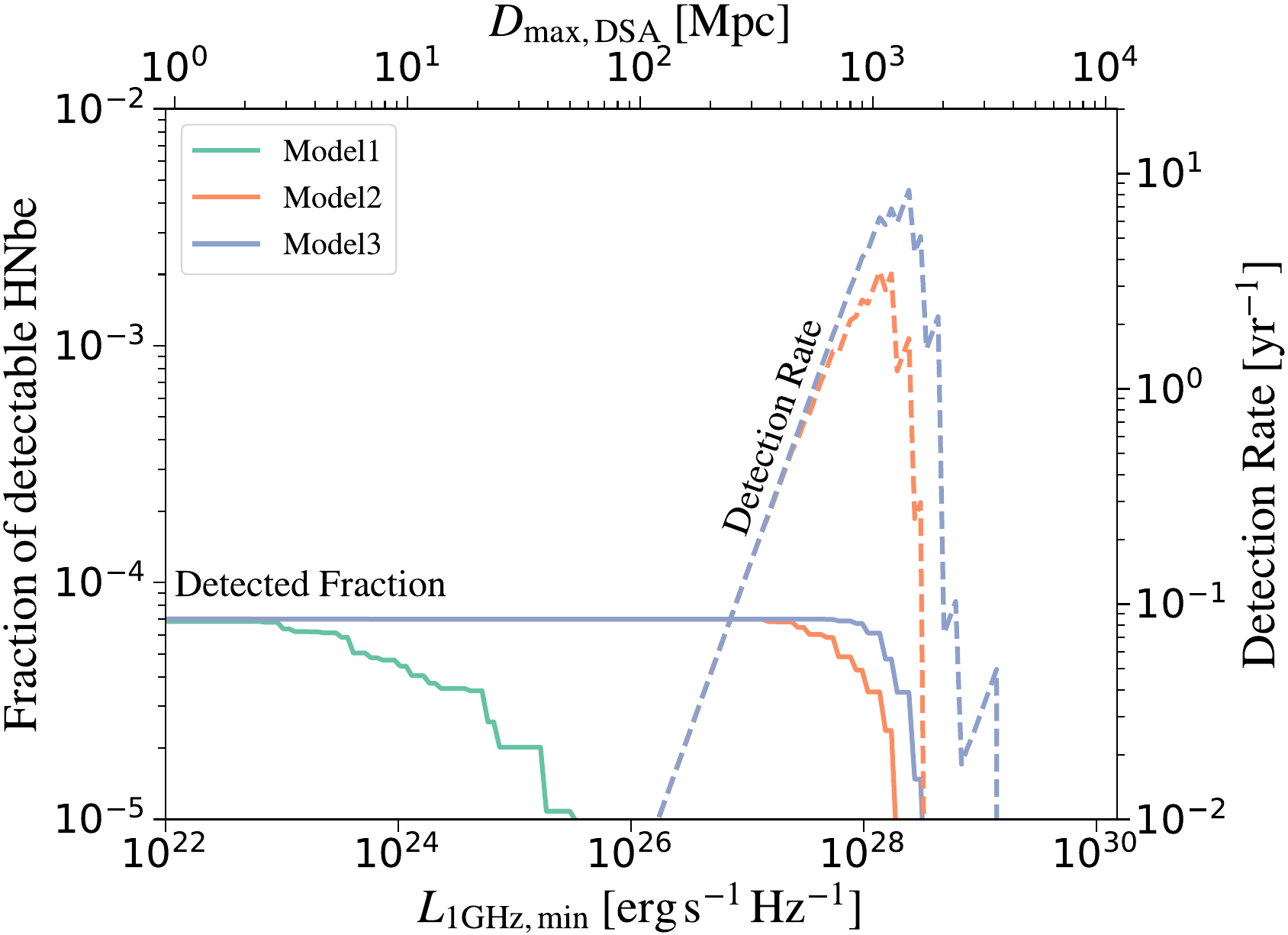}{0.49\textwidth}{(a) Ibc SN progenitors}
    \fig{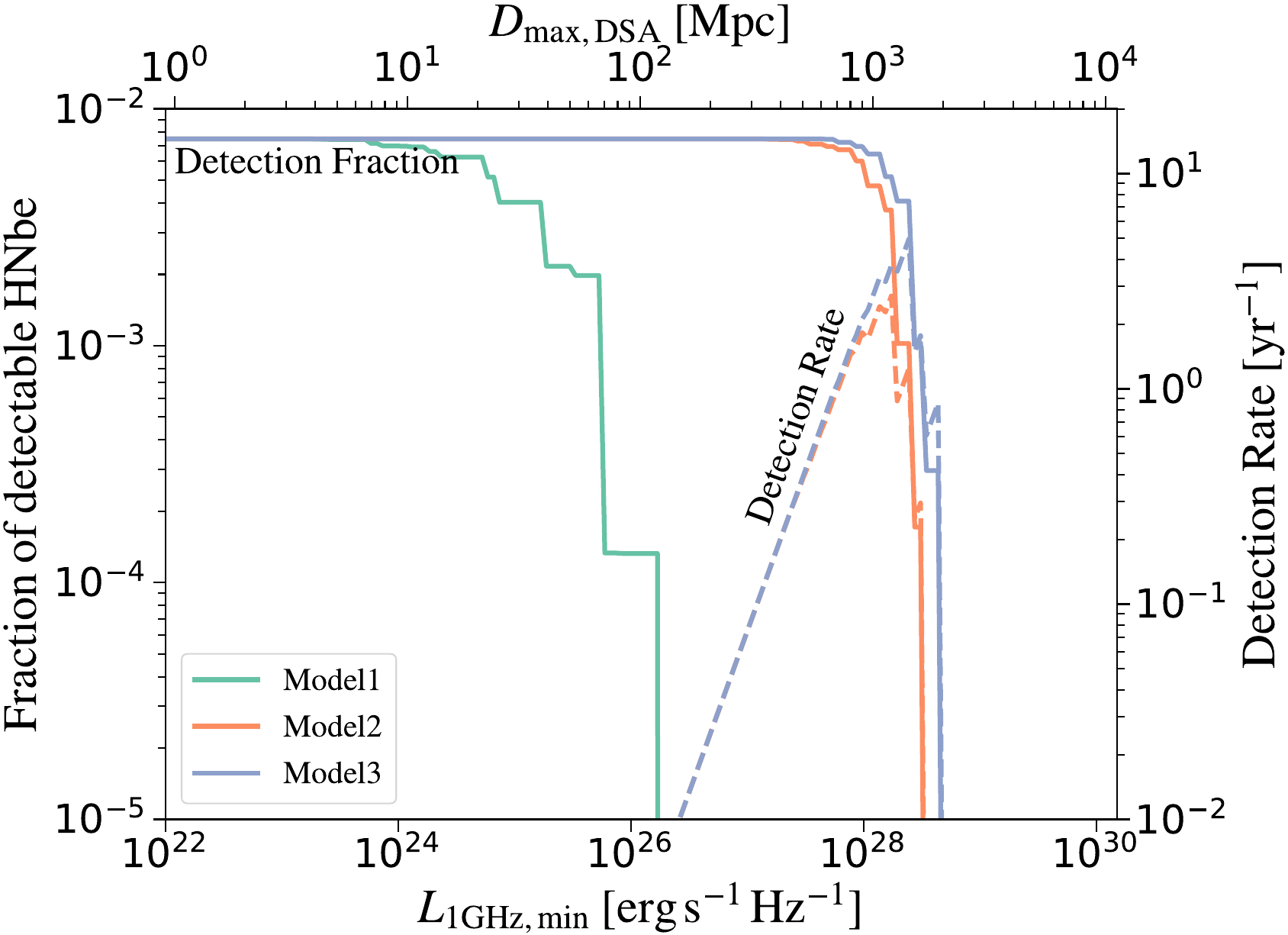}{0.49\textwidth}{(b) LFBOT progenitors}
}
\caption{
Fraction of transient progenitors that harbor HNb emission detectable by DSA-2000 ($F_\mathrm{lim} = 10\,\mu$Jy at 1~GHz). The bottom axis $L_\mathrm{1GHz,min}$ is the luminosity threshold: moving to the right counts only the brighter (and rarer) tail of the HNb population, while the top axis $D_\mathrm{max}^\mathrm{DSA}$ shows the corresponding distance out to which DSA-2000 can detect a source of exactly that luminosity---so larger $L_\mathrm{1GHz,min}$ means a larger survey volume but a smaller fraction of the HNb population detectable coincident with a given transient. The left axis gives this detectable fraction relative to the observed volumetric rate of each transient class (solid lines); the right axis gives the resulting annual detection rate [yr$^{-1}$], which peaks where the two effects balance (dashed lines). Three colors correspond to three jet models. (a) Fraction relative to the observed Type~Ibc SN rate ($\mathcal{R}_\mathrm{Ibc} \sim 2\times10^4$~Gpc$^{-3}$~yr$^{-1}$; \citealt{Frohmaier+21}), including all MT channels with $\bar{t}_\mathrm{delay} < 1$. (b) Fraction relative to the observed LFBOT rate ($\mathcal{R}_\mathrm{LFBOT} \sim 10^2$~Gpc$^{-3}$~yr$^{-1}$; \citealt{Ho_2023ApJ_949_120}), restricted to CE systems with $\bar{t}_\mathrm{delay} < 1$ and $t_\mathrm{delay} < 10^4$~yr.
}
\label{fig:det_fraction}
\end{figure*}

This implies that precursor HNbe emission should be rare among observed stripped-envelope SNe. By comparing the active fraction to the observed volumetric rate of Type~Ibc SNe, $\mathcal{R}_\mathrm{Ibc} \sim 2\times10^4$~Gpc$^{-3}$~yr$^{-1}$ \citep{Frohmaier+21}, we estimate that only $\sim 0.01\%$ of stripped-envelope SNe may exhibit observable HNbe emission at their progenitor locations. This estimate is supported by the detectability estimate shown in Fig.~\ref{fig:det_fraction}(a), where we compute the detectable fraction and annual detection rate for a next-generation radio survey with DSA-2000 sensitivity.

We also consider HNb emission as a possible precursor to LFBOTs \citep[e.g.,][]{Ho_2023ApJ_949_120}. Proposed LFBOT scenarios include mergers between compact objects and stellar cores following CE evolution \citep{Metzger22,Klencki&Metzger26}, as well as tidal disruption of a companion by a newly formed compact object following a supernova kick \citep{Tsuna&Lu25}. These scenarios motivate searching for radio emission powered by binary interaction preceding the transient. As an illustrative proxy for an explosion following recent envelope loss, we select systems in which the second supernova occurs within $10^4$ years after the CE phase. This selection favors systems with relatively little time for their HNb emission to fade, enhancing the detectable precursor fraction relative to the broader stripped-envelope SN population. As shown in Fig.~\ref{fig:det_fraction}(b), the predicted rate of precursors detectable by DSA-2000 corresponds to up to $\sim 1\%$ of the observed LFBOT rate, $\mathcal{R}_\mathrm{LFBOT} = 10^2$~Gpc$^{-3}$~yr$^{-1}$ \citep{Ho_2023ApJ_949_120}. This fraction is roughly two orders of magnitude higher than for Type~Ibc SNe, although the much lower volumetric rate of LFBOTs limits the expected number of associations. The estimate depends on our adopted progenitor selection, which does not explicitly model the merger or tidal-disruption outcomes invoked in these scenarios.

Taken together, these results suggest that detecting HNbe emission as a precursor to binary-driven transients will be difficult with current samples. The expected fractions, $\sim 0.01\%$ for stripped-envelope SNe and $\sim 1\%$ for LFBOT progenitors, are small enough that such associations are unlikely to be represented in the presently known populations in significant numbers. Nevertheless, the situation may change in the era of next-generation time-domain surveys. Rubin Observatory/LSST \citep{LSST+09, LSST_2019_ApJ_873_111} or {\it Roman} is expected to dramatically increase the discovery rate of energetic optical transients, while highly sensitive wide-area radio facilities such as DSA-2000 will be capable of probing the brightest HNb systems over cosmologically significant volumes. For the most luminous sources, Fig.~\ref{fig:det_fraction} implies that DSA-2000 can access distances of order $\sim 1$~Gpc for the most luminous radio sources, opening the possibility that at least a few SN--HNb associations may eventually be discovered. Although challenging, such detections would provide a qualitatively new observational link between pre-transient binary interaction and the final explosive outcome, and therefore offer strong motivation for coordinated optical and radio surveys in the coming decade.

\section{Conclusions} \label{sec:conclusions}

We have developed quantitative predictions for the population properties of hyper-Eddington MT binary systems that produce compact radio nebulae, i.e., hypernebulae \citep{Sridhar2022ApJ}. By combining binary population synthesis modeling with a semi-analytic synchrotron emission model, we evaluated the occurrence rates, physical characteristics, and observational signatures of HNbe across cosmic time.

We find that hyper-Eddington MT systems are formed in the local universe at a volumetric rate of $\mathcal{R} \sim 10^{3}~\mathrm{Gpc^{-3}\,yr^{-1}}$, and exhibit a wide range of MT rates of $\dot{m} \sim 10^{2-7}$ and durations of $10^{0-6}$ years.  These systems include both donor stars undergoing stable MT via RLOF, as well as a brief but extremely high $\dot{M}$ phase immediately preceding unstable MT leading to a CE phase.

Hyper-Eddington accretion onto the CO launches mildly relativistic jets ($v_{\rm j} \sim 0.1$–$0.3c$) and powerful disk winds that inflate long-lived synchrotron nebulae.  For a range of assumed jet velocities: $v_{\rm j}/c = 0.15$, $0.2$, $0.25$ and electron heating efficiencies, we find peak luminosities spanning a wide range up to $L_{\nu,\mathrm{peak}} \sim 10^{28}$–$10^{30}\,\mathrm{erg\,s^{-1}\,Hz^{-1}}$ at GHz frequencies. Since the light-curve duration is mainly controlled by the duration of the long-lasting MT phase, most HNbe are detected as steady radio sources. Given the sensitivity limits of new-generation radio surveys such as VLASS, we estimate that $\sim 10^{2-5}$ HNbe could be detectable for \textsc{Model1}--\textsc{Model2}, whereas \textsc{Model3} predicts more HNbe than the total number of VLASS sources and likely overestimates their luminosities. However, the precise number is extremely sensitive to the assumed jet velocity and electron heating efficiency, which remain theoretically uncertain.

The donor-star properties provide further insight into the evolutionary channels leading to HNbe. Since the binaries driving such high MT must satisfy the Roche lobe, HNbe donors prefer post-main-sequence stars. We find that the most common evolutionary pathways of donors are HG$\to$HeMS in RLOF systems and FGB$\to$HeMS in CE systems. In particular, the most luminous surviving donors such as HeHG stars may be observable in deep imaging with space telescopes such as \textit{Roman} or \textit{JWST}.

A small subset of the detected nebulae (tens of sources in \textsc{Model2}) is expected to be close enough to have angular diameters of $\gtrsim 10$ mas, making them potentially resolvable by VLBI.  These spatially extended sources, along with their predicted angular sizes and morphologies, offer a potential tool for identifying HNbe in nearby galaxies. In addition, the binaries may receive natal kicks during the formation of the compact object, typically displacing them by $\sim 10$ arcsec from their birth sites for nearby ($\lesssim100$ Mpc) sources. This implies that a subset of HNbe may be found far from star-forming regions or galactic centers, providing a further observational signature distinct from other radio populations such as AGN or supernova remnants.

The predicted HNbe properties: persistent luminous radio emission, long-lived accretion phases, and luminous surviving donors, offer valuable context for understanding several classes of unusual transients. In particular, they provide striking similarities to the environments inferred for some FRBs. Indeed, the recent \textit{JWST} possible detection of a luminous near-infrared source coincident with FRB 20250316A \citep{Blanchard_2025_ApJ_989_L49}, potentially consistent with the luminous end of HeHG/HeGB stars, may be compatible with this scenario, despite the non-detection of radio emission at its location. Future radio and multi-wavelength surveys will be essential to test this scenario and to reveal the broader role of hyper-Eddington MT in transient astrophysics and binary evolution.

Our models for HNb emission and binary star evolution make several simplifying assumptions. The predicted detectable population is sensitive to uncertain microphysical parameters and to the treatment of binary interactions, particularly common-envelope evolution. Our exclusion of merging systems also omits a potentially important HNb formation channel: binaries that ultimately coalesce may undergo substantial MT and launch accretion-powered outflows before merger. Their contribution depends on the duration of this accretion phase and whether the outflows can escape the surrounding envelope. Consequently, the small CE contribution found here does not establish that unstable MT contributes little to the full HNb population.

Nevertheless, we expect the qualitative features highlighted here, namely the existence of a luminous long-lived radio population, its broad luminosity range, and the utility of spatial extent, offsets, and donor-star counterparts as identifying signatures, to be more robust than any single numerical estimate. More generally, the methodology laid out in this paper provides a proof of principle for connecting binary population synthesis calculations to non-transient radio source populations in wide-field surveys. Future progress should come from both theoretical and observational advances. On the theoretical side, continued improvements in population-synthesis tools, including recent updates to \textsc{COMPAS} that implement more modern treatments of massive-star winds \citep{Merritt2025arXiv250717052M}, together with a better understanding of electron energization microphysics in magnetized shocks and light-curve calculations that incorporate non-thermal as well as thermal electron populations, should place these predictions on a firmer quantitative footing. On the observational side, deeper and higher-resolution radio surveys, together with multi-wavelength follow-up, should provide much stronger tests of the HNb scenario.

\section*{Acknowledgments}

KU acknowledges financial support from Grant-in-Aid for the Japan Society for the Promotion of Science (JSPS) Fellows (22J22705) and JSPS Overseas Research Fellowship (No. 202560419). N.S. acknowledges support from the Simons Foundation (grant MP-SCMPS-00001470) and the National Aeronautics and Space Administration (NASA grant 80NSSC26K0946).  BDM acknowledges support from NASA (grant 80NSSC24K0408), the National Science Foundation (grant AST-2406637), and the Simons Foundation (grant 727700). FSB acknowledges support from NSF 22-624 Astronomy and Astrophysics Research Grants under award number 2606407.  The Flatiron Institute is supported by the Simons Foundation. KM acknowledges support from the Japan Society for the Promotion of Science (JSPS) KAKENHI grant (JP24KK0070, JP24H01810, and JP23H04894).

\section*{Data Availability}
The \textsc{COMPAS} population synthesis data used in this work are publicly available from \citet{vanSon+22}. The hypernebula light-curve model grid will be made publicly available on Zenodo upon publication.

\software{
          \textsc{COMPAS} \citep{Riley2022ApJS}
          astropy \citep{astropy:2013, astropy:2018, astropy:2022},
          NumPy \citep{Numpy},
          Matplotlib \citep{Matplotlib}
          }

\bibliography{manuscript}{}
\bibliographystyle{aasjournal}

\appendix

\section{Hypernebula Model Details} \label{sec:radiolc}
Here we describe in greater detail our implementation of the \citetalias{Sridhar2022ApJ} HNb light curve model.

The kinetic luminosity of the jet,
\begin{equation}
    L_{\rm j} = \eta L_{\rm w} = \eta \frac{1}{2} \dot{M}_{\rm w} v_{\rm w}^2,
    \label{eq:Lj}
\end{equation}
is assumed to scale with the kinetic luminosity of the wind $L_{\rm w} = (1/2)\dot{M}_{\rm w}v_{\rm w}^{2}$, where $\eta$ is a free parameter (see Sec.~\ref{sec:jet_efficiency}) and $\dot{M}_{\rm w}$ and $v_{\rm w}$ are the wind mass-loss rate and velocity, respectively.  We take $\dot{M}_{\rm w} = \dot{M}$ because the powerful outflows that accompany hyper-Eddington accretion are expected to carry away the vast majority of the transferred mass.

The distribution of electrons with energy $E = \gamma m_e c^{2}$ in the nebula, $N_{\gamma}$, evolves with time $t$ since the onset of the MT phase according to:
\begin{align}
    \frac{dN_{\gamma}}{dt} = \frac{\partial}{\partial t}N_{\gamma} - \frac{\partial}{\partial \gamma}(\dot{\gamma} N_{\gamma}) - N_{\gamma} \frac{d}{dt} \ln V_{n}.
    \label{eq:dNdt}
\end{align}
The first term here accounts for the injection of relativistic electrons into the nebula at the jet termination shock. We assume the electrons possess a thermal Maxwellian energy distribution characterized by
\begin{equation}
    \bar{\gamma} \approx \frac{1}{2} \varepsilon_{\mathrm{e}}\frac{m_{\mathrm{p}}}{m_{\mathrm{e}}}\frac{v_{\rm j}^2}{c^2},
    \label{eq:Ngamma}
\end{equation}
injected at a total rate $\dot{N}_{\rm e} = L_{\rm j}/[2\bar{\gamma}m_{\rm e}c^{2}(1+\sigma_{\rm j})]$, where $v_{\rm j}$ is the jet speed, $m_{\mathrm{p}}/m_{\rm e}$ is the proton-to-electron mass ratio, and $\varepsilon_{\mathrm{e}}$ is the electron heating efficiency at the shock. The second and third terms in Eq.~\eqref{eq:dNdt} correspond to radiative and adiabatic expansion losses, respectively, where $V_{\rm n} = (4\pi/3)R_{\rm n}^{3}$ is the nebula volume of radius $R_{\rm n}$. The latter is set by a balance between the ram pressure of the jet and the pressure of the surrounding shell of mass driven into the CSM by the disk winds. The radius of the shell, $R_{\rm sh}$, evolves through an initial free-expansion phase, $R_{\rm sh}=v_{\rm w}t$, followed by an energy-driven (adiabatic) expansion once the mass of swept-up CSM exceeds the injected wind mass,
\begin{equation}
R_{\rm sh}\simeq 0.8\left(\frac{E(t)\,t^{2}}{\rho_{\rm csm}}\right)^{1/5},
\end{equation}
where $E(t)$ is the energy injected by the wind up to time $t$, so that $R_{\rm sh}\propto t^{3/5}$, while the wind is active ($E = L_{\rm w}t$; \citealt{Weaver+77}) and $R_{\rm sh}\propto t^{2/5}$ after it turns off (see \citetalias{Sridhar2022ApJ} for details).

The magnetic energy of the nebula, $E_{\mathrm{B}}$, and hence its magnetic field strength $B_{\rm n}$, evolves according to:
\begin{equation}
    \frac{dE_{\mathrm{B}}}{dt} = \left(\frac{\sigma_{\rm j}}{1+\sigma_{\rm j}}\right) L_{\rm j} - \frac{\dot{R_{\mathrm{n}}}}{R_{\mathrm{n}}} E_{\mathrm{B_{\mathrm{n}}}};
    \quad E_{\mathrm{B}} = \left(\frac{B_{\mathrm{n}}^2}{8\pi}\right)V_{\mathrm{n}},
\end{equation}
where the source term is proportional the magnetization of the jet $\sigma_{\rm j}$, i.e. the fraction of $L_{\rm j}$ carried by Poynting flux.

Given the distribution of electrons ($N_{\gamma}$) and magnetic field strength of the nebula $B_{\mathrm{n}}$, the radio synchrotron luminosity ($L_{\nu}$) is calculated as follows:
\begin{equation}
    L_{\nu} = 4\pi^{2} R_{\mathrm{n}}^2 \frac{j_{\nu}}{\alpha_{\nu}}\left\{1 - \exp(- \alpha_{\nu} R_{\mathrm{n}}) \right\},
\end{equation}
where $j_{\nu}$ and $\alpha_{\nu}$ are the emissivity and absorption coefficient, respectively, defined according to:
\begin{equation}
j_{\nu}=\int \frac{N_{\gamma} P_{\nu}(\gamma)}{4 \pi} d \gamma;
\quad 
\alpha_{\nu}=-\int \frac{\gamma^2 P_{\nu}(\gamma)}{8 \pi m_{\mathrm{e}} \nu^2} \frac{\partial}{\partial \gamma}\left[\frac{N_{\gamma}}{\gamma^2}\right] d \gamma,
\end{equation}
where $m_{\mathrm{e}}$ is the electron mass, $\gamma$ is the electron Lorentz factor, and
\begin{equation}
P_{\nu}=\frac{2 e^3 B_{\mathrm{n}}}{\sqrt{3} m_{\mathrm{e}} c^2} F\left(\frac{\nu}{\nu_{\mathrm{syn}}}\right);
\quad 
F(x) \equiv x \int_x^{\infty} K_{5 / 3}(y) dy,
\end{equation}
is the spectral power of a synchrotron-emitting electron, where $e$ is the elementary charge, $c$ is the light speed, $K_{5/3}$ is the modified Bessel function of order $5/3$, and
\begin{equation}
    \nu_{\mathrm{syn}} = \frac{eB_{\mathrm{n}}}{2\pi m_{\mathrm{e}}c}\gamma^{2}
\end{equation}
is the characteristic synchrotron frequency. 

\subsection{Jet Efficiency}
\label{sec:jet_efficiency}

Here we motivate the jet efficiency used in our models following recent simulation results of hyper-Eddington accretion. Consider a compact object of mass $M_{\rm CO}$ is being fed through a disk starting from some outer radius $R_{\rm d}$ at a rate $\dot{M}$ from the binary star.  The radius of the disk $R_{\rm d}(q,M_{\rm bin})$ is set by the angular momentum of matter through the $L_1$ Lagrange point and can be related to the binary period and mass-ratio $q$ according to (e.g. \citealt{Frank+02}; their Eq.~4.21):
\begin{equation}
R_{\rm d} = 4 R_{\odot} P_{\rm d}^{2/3}(1+q)^{4/3}\left[0.5-0.23{\rm log_{10}}(q)\right]^{4},
\end{equation}
where $P_{\rm d}$ is the orbital period in days.  Photons are trapped in the disk out to a radius \citep{Begelman79},
\begin{equation}
R_{\rm t} \simeq 10R_{\rm g}\left(\frac{\dot{M}}{\dot{M}_{\rm Edd}}\right) \approx 2R_{\odot}\left(\frac{M_{\rm CO}}{10M_{\odot}}\right)\left(\frac{\dot{M}}{10^{4}\dot{M}_{\rm Edd}}\right),
\end{equation}
where again we define $\dot{M}_{\rm Edd} \equiv L_{\rm Edd}/(0.1 c^{2})$, $R_{\rm g} \equiv GM_{\rm CO}/c^{2}$.  As a result of hyper-Eddington disk outflows, the mass-accretion rate which reaches the central CO is reduced from the feeding rate according to \citep{Blandford&Begelman99}
\begin{equation}
\dot{M}_{\rm CO} = \dot{M}\left(\frac{R_{\rm out}}{R_{\rm ISCO}}\right)^{-p},
\label{eq:Mdot_CO}
\end{equation}
where $p \simeq 0.7$ \citep{Yuan&Narayan14}, $R_{\rm ISCO} = 6 R_{\rm g}$ is the radius of the innermost stable circular orbit for a non-spinning black hole, and $R_{\rm out}$ represents the minimum of the outer radius of the disk and the trapping radius, i.e.
\begin{equation}
R_{\rm out} = {\rm min}\left[R_{\rm t},R_{\rm d}\right].
\end{equation}
Except for very short orbital periods and extremely high accretion rates, we generally have $R_{\rm t} \lesssim R_{\rm d}$.  Disk outflows are launched from a range of radii $R_{\rm ISCO} \ll r \ll R_{\rm out}$, with an average velocity \citep{Margalit&Metzger16}:
\begin{equation}
v_w \approx \sqrt{\frac{p}{1-p}\frac{GM_{\rm CO}}{R_{\rm ISCO}}\left(\frac{R_{\rm ISCO}}{R_{\rm out}}\right)^{p}} \underset{p = 0.7}\approx 0.03 c\,\, M_{\rm CO,10}^{0.35}\left(\frac{R_{\rm out}}{R_{\odot}}\right)^{-0.35}.
\label{eq:vw}
\end{equation}
According to recent radiation hydrodynamic simulations of super-Eddington accretion by \citet{Yoshioka+24}, the kinetic power of the fast outflow launched from the inner accretion disk scales with the CO accretion rate. Although Eq.~\eqref{eq:Mdot_CO} predicts a strong reduction of $\dot{M}_{\rm CO}$ for $p \simeq 0.7$, a fraction $\sim 0.3$--$0.7$ of the injected mass reaches the black hole in their simulations (their Table~2); we therefore approximate $\dot{M}_{\rm CO} \approx 0.5\dot{M}$, such that
\begin{equation}
\frac{L_{\rm j,Y}}{L_{\rm Edd}} \simeq \left(\frac{\dot{M}_{\rm CO}}{50\dot{M}_{\rm Edd}}\right)^{0.7} \approx 0.04\left(\frac{\dot{M}}{\dot{M}_{\rm Edd}}\right)^{0.7}.
\end{equation}
To match our assumed prescription for the jet power (Eq.~\eqref{eq:Lj}),
\begin{equation}
\frac{L_{\rm j}}{L_{\rm Edd}} = \frac{(\eta/2)\dot{M}v_{\rm w}^{2}}{L_{\rm Edd}} = 5\eta \left(\frac{\dot{M}}{\dot{M}_{\rm Edd}}\right)\left(\frac{v_{\rm w}}{c}\right)^{2}
\end{equation}
to the results of \citet{Yoshioka+24}, we thus require an effective jet efficiency:
\begin{eqnarray}
\eta_{\rm Y} \approx 0.6\left(\frac{\dot{M}}{10^{4}\dot{M}_{\rm Edd}}\right)^{-0.3}\left(\frac{v_{\rm w}}{0.03c}\right)^{-2} 
\approx 0.9\left(\frac{\dot{M}}{10^{4}\dot{M}_{\rm Edd}}\right)^{0.4}\left(\frac{R_{\rm out}}{R_{\rm t}}\right)^{0.7},
\end{eqnarray}
where we have used Eq.~\eqref{eq:vw} in the second line, normalizing $R_{\rm out}$ to the trapping radius. This normalization and its weak dependence on $\dot{M}$, i.e. $\eta \sim 0.1-1$, as well as uncertainties associated with extrapolating the results of \citet{Yoshioka+24} to higher $\dot{M}$ than covered by their simulations, motivate our adoption of a fixed jet efficiency $\eta \sim 0.1$.

\section{Hypernebula Model for SS433} \label{sec:ss433}

\begin{figure*}
\gridline{
    \fig{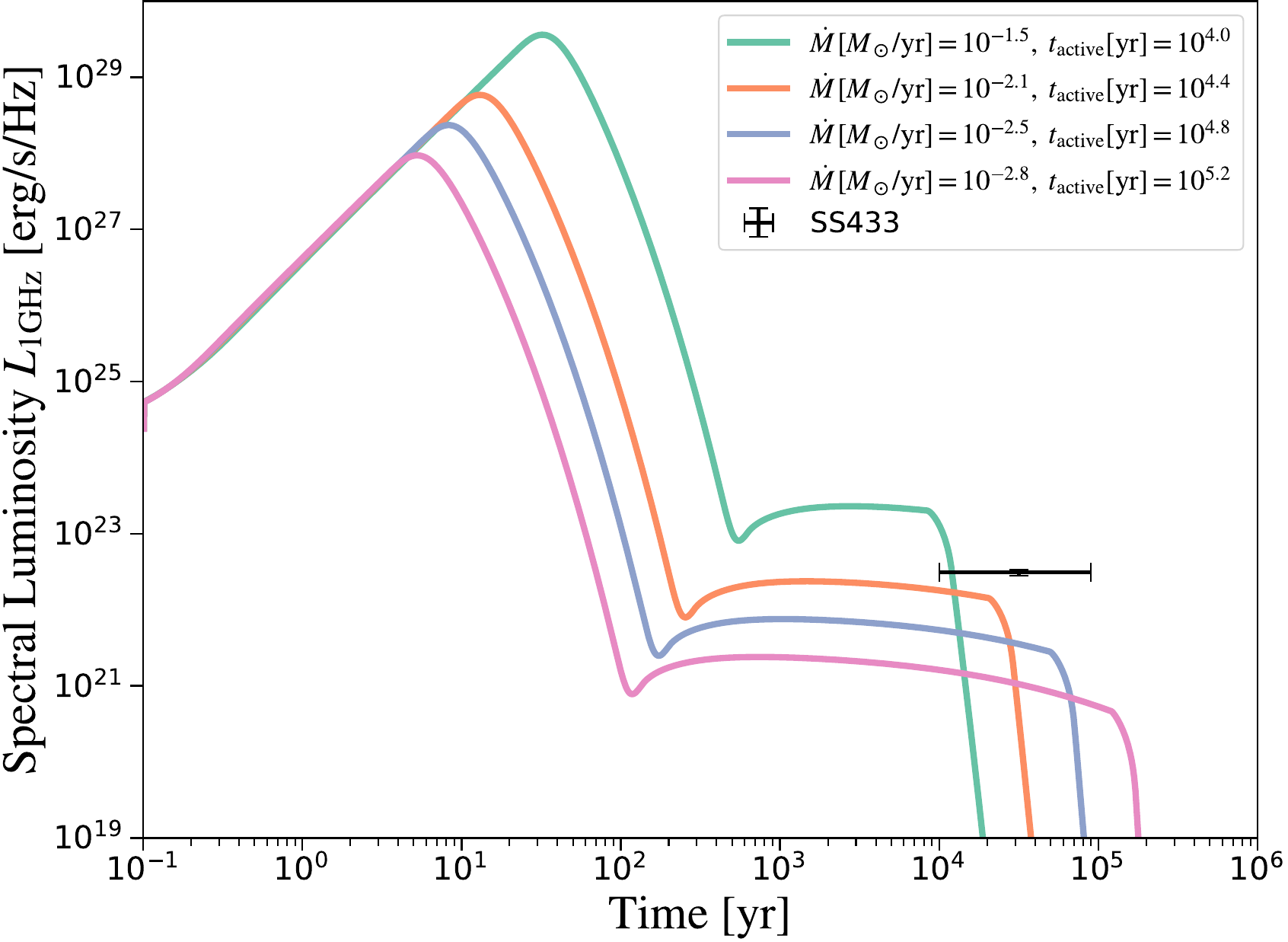}{0.49\textwidth}{(a) 1GHz light curve for the adopted parameters}
    \fig{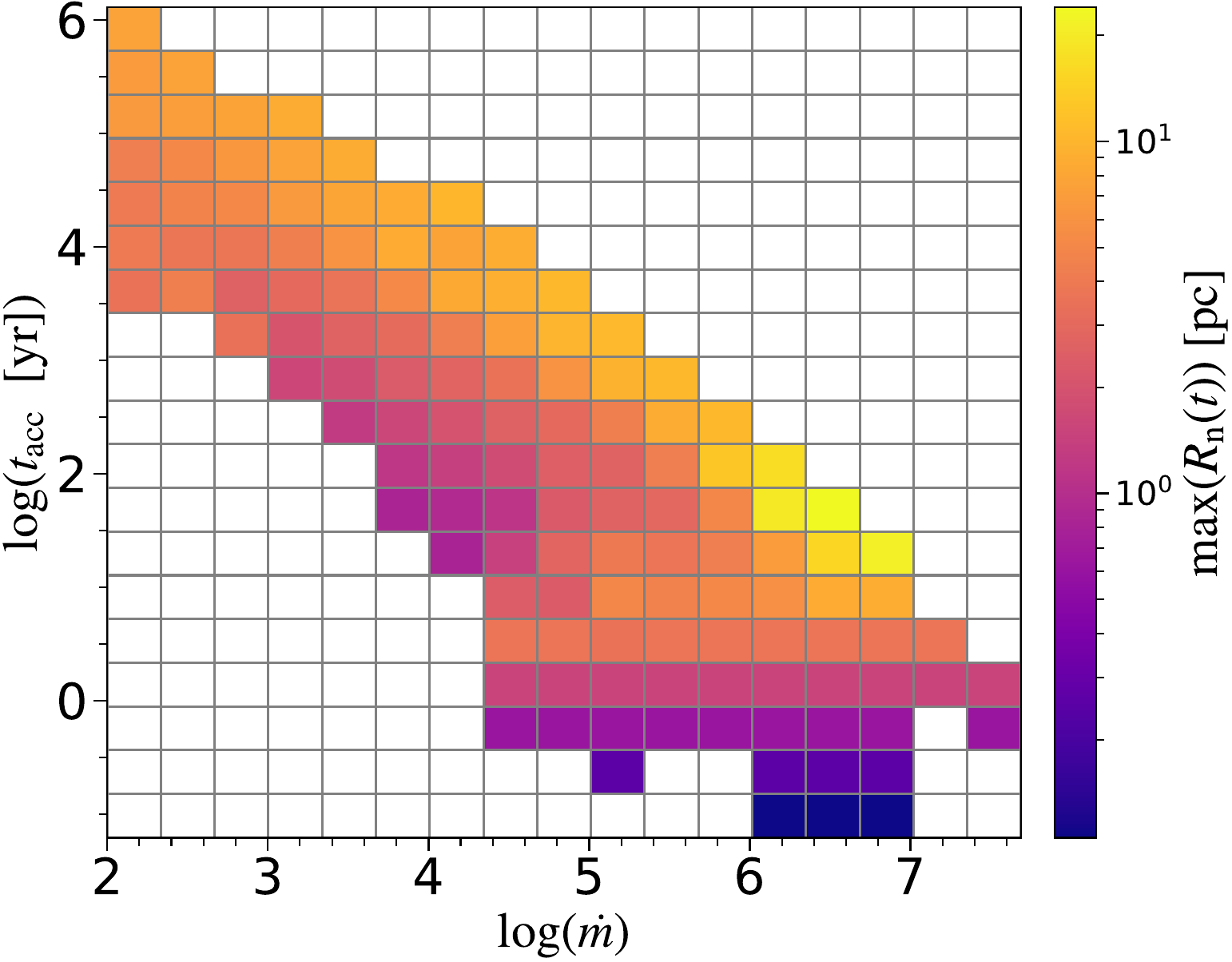}{0.49\textwidth}{(b) Maximum nebula size}
}
\gridline{
    \fig{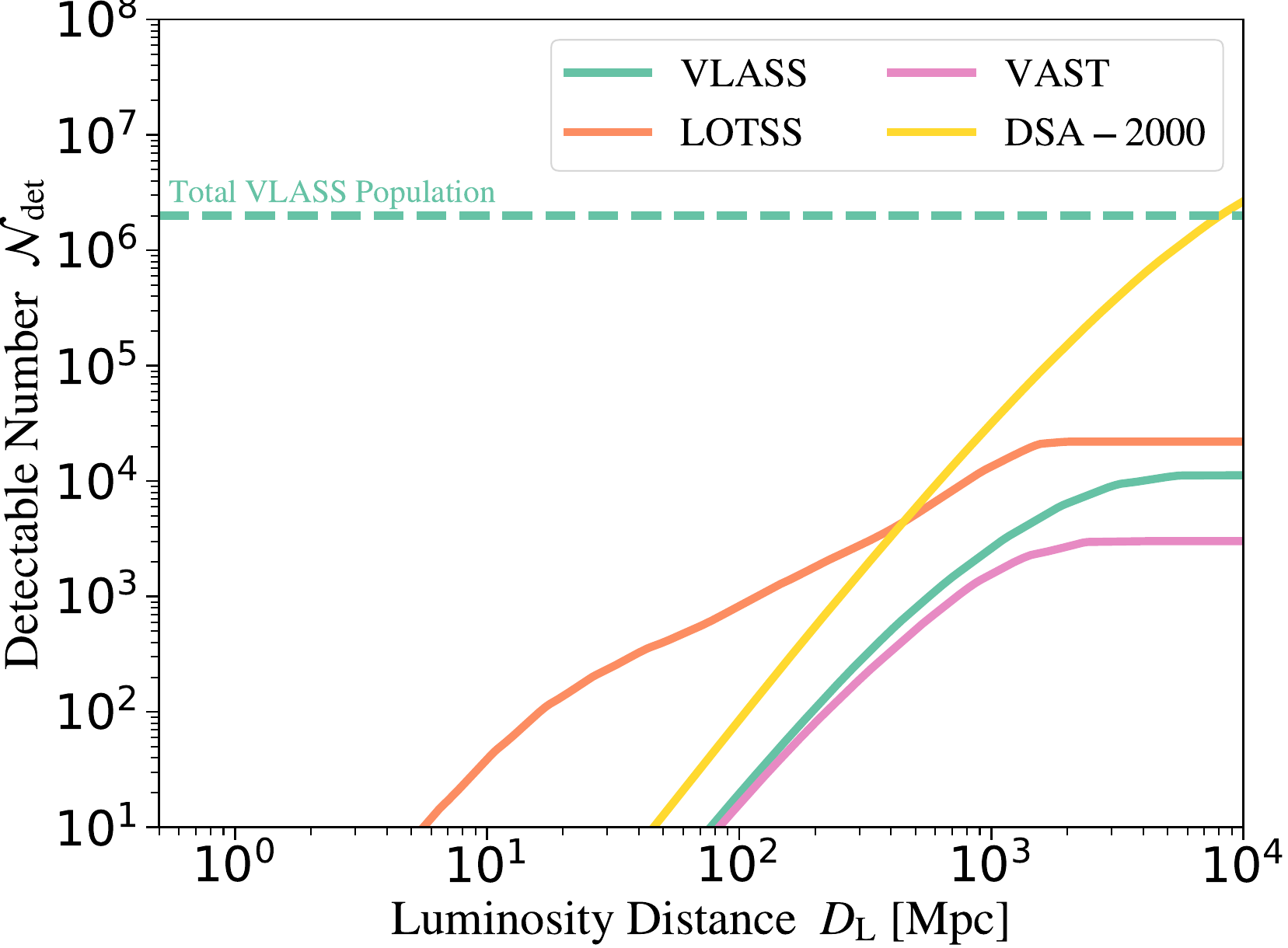}{0.49\textwidth}{(c) Cumulative number of HNb sources}
    \fig{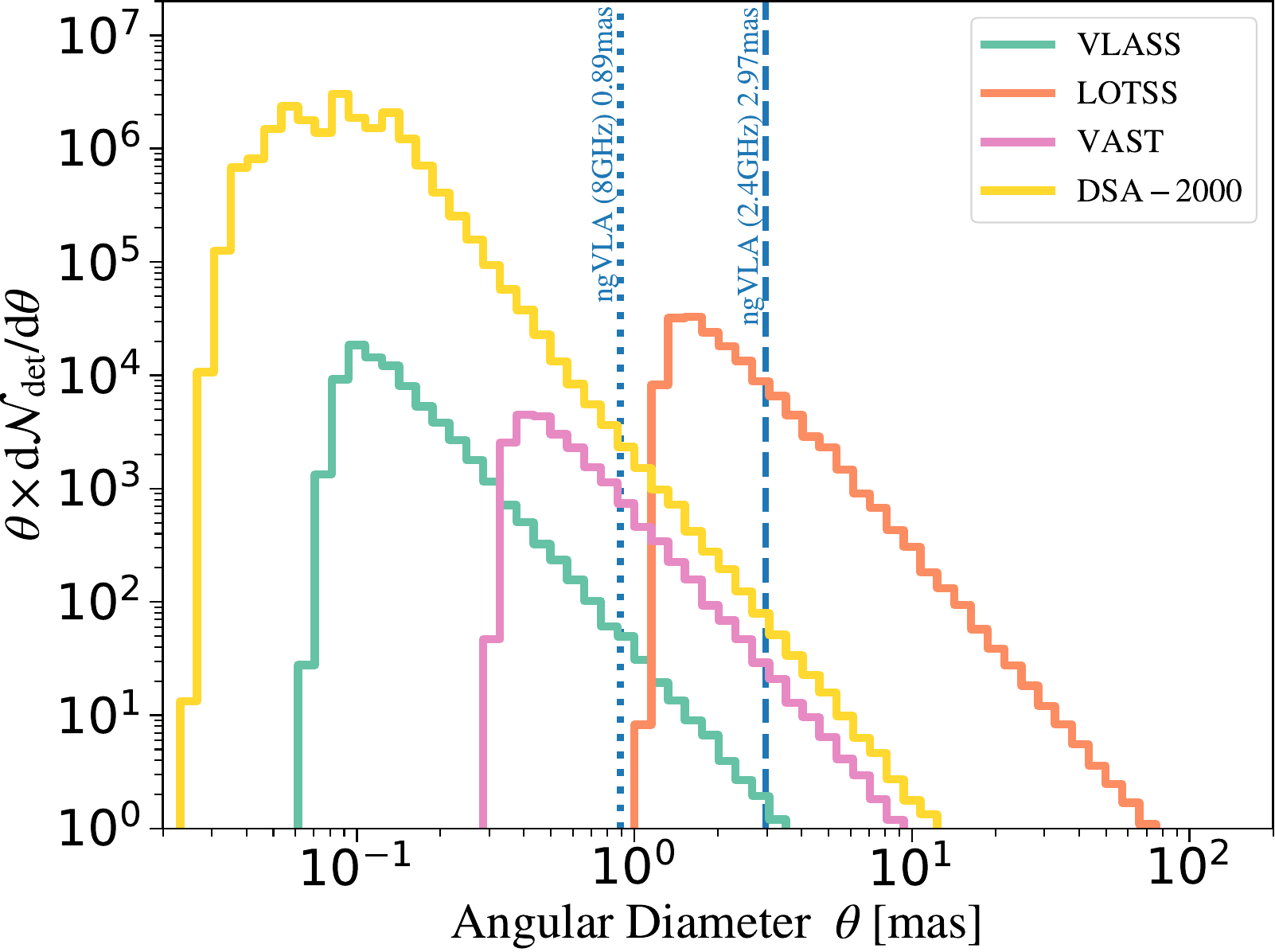}{0.49\textwidth}{(d) Distribution of angular diameter size}
}
\caption{
(a) HNbe light curves for various MT systems as marked, for parameters described in Appendix \ref{sec:ss433}. The observed lifetime and luminosity of SS433 are shown for comparison by the cross marker.
(b) Maximum nebula size for the adopted parameters. Some models reach sizes of a few $\times 10$ pc.
(c) Same as Fig.~\ref{fig:cumulative_models}, but for the SS433-motivated parameter set.
(d) Same as Fig.~\ref{fig:dNdet_theta}, but for the SS433-motivated parameter set.
}
\label{fig:appendix_ss433}
\end{figure*}

Here, we present a model for SS433. To reproduce its typical physical scale for high energy emission ($\sim 20$ pc at the jet break; e.g., \citealt{Aharonian_2024_Science_383_402}) and radio luminosity ($\sim$ a few $\times 10^{22}\,\mathrm{erg\,s^{-1}\,Hz^{-1}}$; e.g., \citealt{Trushkin_2003_BSAO_56_57}), we adopt the following parameter set: $(v_{\mathrm{j}}/c, v_{\mathrm{w}}/c, \varepsilon_{\mathrm{e}}, \eta, \sigma_{\mathrm{j}}, n_{\mathrm{csm}}) = (0.26, 0.1, 0.1, 0.5, 0.1, 0.01)$. The corresponding results are shown in Fig.~\ref{fig:appendix_ss433}. Note that this comparison is intended only as an order-of-magnitude consistency check. Because the model contains substantial uncertainties in both the physical assumptions and parameter selections, the adopted parameter set is not unique, and an exact match to all observed properties of SS433 is not required. 
Also, we note that \citet{Wang_2026_ApJ_1003_118} reports that recent VLBA observations of extragalactic ULXs do not detect mas-scale compact radio cores. However, these non-detections do not rule out SS433-like jets and associated pc-scale extended structures in ULXs, because compact cores driven by SS433-like jets would be too faint at the GHz frequency to detect at extragalactic distances. In addition, VLBI observations with mas-scale resolution are not sensitive to diffuse pc-scale structures.

\section{Star Formation Rate Density} \label{sec:sfrd}

We use the \citet{Madau2014ARAA} fitting function for the total (metallicity-independent) cosmic star formation history:
\begin{equation}
    {\mathrm {SFRD}} = \frac{d^2 M}{dtdV} = a \frac{(1+z)^b}{1 + \left(\frac{1+z}{c}\right)^{d}} {\rm ~[M_{\odot}/yr/Mpc^{3}]},
\end{equation}
where we adopt the fitting parameters provided by \citet{Neijssel2019MNRAS}; $(a, b, c, d) = (0.01, 2.77, 2.9, 4.7)$.

For the metallicity distribution, we assume a log skew normal distribution as follows:
\begin{align}
    \frac{dP}{dZ} (Z,z) &= \frac{1}{Z}\frac{dP}{d\ln Z} = \frac{1}{Z} \times \frac{2}{\sigma(z)} \phi\left(\frac{\ln Z-\xi(z)}{\sigma(z)}\right) \Phi\left(\alpha \frac{\ln Z-\xi(z)}{\sigma}\right),
\end{align}
where $\phi(x)$ and $\Phi(x)$ are the standard normal distribution and the cumulative of the normal distribution;
\begin{equation}
    \phi(x) = \frac{1}{\sqrt{2\pi}} \exp \left( -\frac{1}{2}x^2 \right); \quad \Phi(x) = \frac{1}{2} \left\{ 1 + \mathrm{erf} \left( \frac{x}{\sqrt{2}} \right) \right\}.
\end{equation}
Here, 
\begin{equation}
    \sigma(z) = \sigma_{0} \cdot 10^{\sigma_{z}\cdot z},
\end{equation}
is the redshift-dependent metallicity-variance distribution, where $\sigma_{0} = \sigma(z\approx 0)$ and $\sigma_{z}$ are redshift scaling parameters. Likewise, the function
\begin{equation}
    \xi(z) = \ln \left( \frac{\mu(z)}{2\Phi(\beta \sigma(z))} \right) - \frac{\sigma(z)^2}{2},
\end{equation}
is the redshift-dependent mean metallicity distribution.  Here, the mean metallicity $\mu(z)$ obeys
\begin{equation}
    \mu(z) =  \mu_{0} \cdot 10^{\mu_{z}\cdot z},
\end{equation}
where $\mu_{0} = \mu(z\approx 0)$ and $\mu_{z}$ is again a redshift-scaling parameter for the mean-metallicity. Finally, the scaling parameter $\beta$ depends on the skewness ($\alpha$) of the log-normal distribution, according to:
\begin{equation}
    \beta = \frac{\alpha}{\sqrt{1 + \alpha^2}}
\end{equation}
In this paper, we adopt the parameter set given by \citet{vanSon+23}: $(\mu_{0}, \mu_{z}, \sigma_{0}, \sigma_{z}, \alpha) = (0.025, -0.048, 1.125, 0.048, -1.767)$.

\end{document}